\documentclass[12pt,a4paper]{article}
\usepackage{graphics,amsmath,amssymb,xcolor,pstool,braket,amsthm,mathtools,comment,caption,multirow,soul,float,cite}

\usepackage[colorlinks = true,
            linkcolor = blue!70!black,
            urlcolor  = red!70!black,
            citecolor = green!55!black,
            anchorcolor = blue!70!black,bookmarks]{hyperref}

\def\be{\begin{equation}}
\def\ee{\end{equation}}
\def\di{\mathrm{d}}

\begin{document}
\begin{center}

{\Large \bf Non-analyticity of the $S$-matrix \\[.3cm] with spontaneously broken Lorentz invariance}  \\[0.7cm]

\large{Paolo Creminelli$^{\,\rm a, \rm b}$, Matteo Delladio$^{\,\rm c}$, Oliver Janssen$^{\,\rm d}$, \\
Alessandro Longo$^{\,\rm e, \rm f}$ and Leonardo Senatore$^{\,\rm c}$}
\\[0.5cm]

\small{
\textit{$^{\rm a}$
ICTP, International Centre for Theoretical Physics, 34151 Trieste, Italy}}
\vspace{.2cm}

\small{
\textit{$^{\rm b}$
IFPU, Institute for Fundamental Physics of the Universe, 34014 Trieste, Italy}}
\vspace{.2cm}

\small{
\textit{$^{\rm c}$ Institut f\"ur Theoretische Physik, ETH Z\"urich, 8093 Z\"urich, Switzerland}}
\vspace{.2cm}

\small{
\textit{$^{\rm d}$ Laboratory for Theoretical Fundamental Physics, EPFL, 1015 Lausanne, Switzerland}}
\vspace{.2cm}

\small{
\textit{$^{\rm e}$ SISSA, International School for Advanced Studies, 34136 Trieste, Italy}}
\vspace{.2cm}

\small{
\textit{$^{\rm f}$ INFN, National Institute for Nuclear Physics, 34127 Trieste, Italy}}
\vspace{.2cm}
\vspace{1cm}
\end{center}

\hrule \vspace{0.3cm}
{\small  \noindent \textbf{Abstract} \noindent \quad
We study the $S$-matrix of Goldstones in the renormalizable theory of a $U(1)$ complex scalar at finite charge, i.e.~in a state that breaks Lorentz invariance. The theory is weakly coupled so that this $S$-matrix exists at all energies.  Unlike the Lorentz invariant case, the resulting $S$-matrix is not analytic in the exchanged (complexified) four-momentum. The non-analyticities stem from the  LSZ reduction formula, as a consequence of the energy-dependent mixing between the radial and Goldstone modes.
\noindent 
\vspace{0.3cm}}
\hrule

\vspace{0.3cm}
\newpage
\tableofcontents
\section{Introduction}
Locality and unitarity impose stringent constraints on the $S$-matrix of consistent effective field theories (EFTs). The derived positivity bounds \cite{Adams:2006sv, Bellazzini:2020cot, Caron-Huot:2020cmc, Tolley:2020gtv, Arkani-Hamed:2020blm} and the general program of $S$-matrix bootstrap \cite{Paulos:2016fap, EliasMiro:2022xaa} are under intense investigation. Nearly all research has been conducted under the assumption of (linearly realized) Lorentz invariance, which is a natural presumption in the context of particle physics. However, Lorentz invariance is spontaneously broken in cosmology (inflation, dark energy, \dots), in condensed matter and also when studying perturbations of localized objects like black holes (see for instance \cite{Delacretaz:2014oxa, Franciolini:2018uyq}). It would be valuable to derive bounds on EFTs without relying on Lorentz invariance.

The simplest approach would be to focus on the $S$-matrix and follow similar steps as in the Lorentz invariant case \cite{Baumann:2015nta,Grall:2021xxm}. However, it is evident that studying the $S$-matrix is not appropriate for a general UV completion, as discussed in \cite{Creminelli:2022onn}. In the presence of Lorentz invariance one can scatter asymptotic states at any energy, so that the $S$-matrix is well-defined both in the UV and in the IR. This does not hold in general when Lorentz invariance is absent: the low energy excitations around a vacuum that spontaneously breaks Lorentz (phonons, inflaton perturbations, quasinormal modes of a black hole, \dots) simply do not exist at high energy, so that one cannot define the $S$-matrix at arbitrary energies. In general one has to look for other quantities which are well-defined both in the UV and in the IR to derive universal bounds: one possibility is to look at conserved currents as was done in \cite{Creminelli:2022onn}.

In this paper we study the $S$-matrix in a model that can be considered the simplest example of spontaneous breaking of Lorentz invariance: the renormalizable theory of a  complex scalar with a $U(1)$ symmetry in a state with finite charge. Boosts are spontaneously broken, while space and time translations are not.\footnote{The unbroken time translations actually correspond to a linear combination of the original time translation and a $U(1)$ transformation \cite{Nicolis:2013lma}.} The model features two modes: a massless Goldstone and the radial massive excitation, whose dispersion relations will be studied in section~\ref{secU1model}. The model is weakly coupled so that, contrary to the general case discussed above, one can study the $S$-matrix of the Goldstones at all energies. Before venturing to make very general claims, we think it is important to have at least one non-trivial example of an $S$-matrix in the absence of Lorentz invariance.

Before calculating the $S$-matrix and evaluating it in various limits of physical interest (section~\ref{secSmatrix} and appendices~\ref{appFeynman} and \ref{appSMatrixLowEnergy}), in section~\ref{secLSZ} (and appendix~\ref{app2LSZ}) we will derive an LSZ reduction formula that connects the $S$-matrix to correlation functions. Here we encounter a striking difference compared to the Lorentz invariant case: the relation between the $S$-matrix and correlation functions involves operators that are non-local in space. This spatial non-locality arises from the mixing of the two modes, which depends on the momentum. The presence of this non-locality spoils the analyticity properties of the $S$-matrix compared to the Lorentz invariant case. We conclude that even in cases where the $S$-matrix exists at all energies one cannot derive positivity bounds using the conventional dispersive arguments.

In section~\ref{secdecay} and appendix~\ref{appdecay} we study the decay rate of Goldstones, while conclusions and future directions are discussed in section~\ref{secconclusions}.

{\bf Note added:} We understand that Hui, Kourkoulou, Nicolis, Podo and Zhou have been working on essentially the same questions and model. We did not exchange results, but we did coordinate the arXiv submission. 

\section{\label{secU1model}The \texorpdfstring{$U(1)$}{U1} model}
\begin{figure}[ht]
    \centering
    \includegraphics[width=0.75\textwidth]{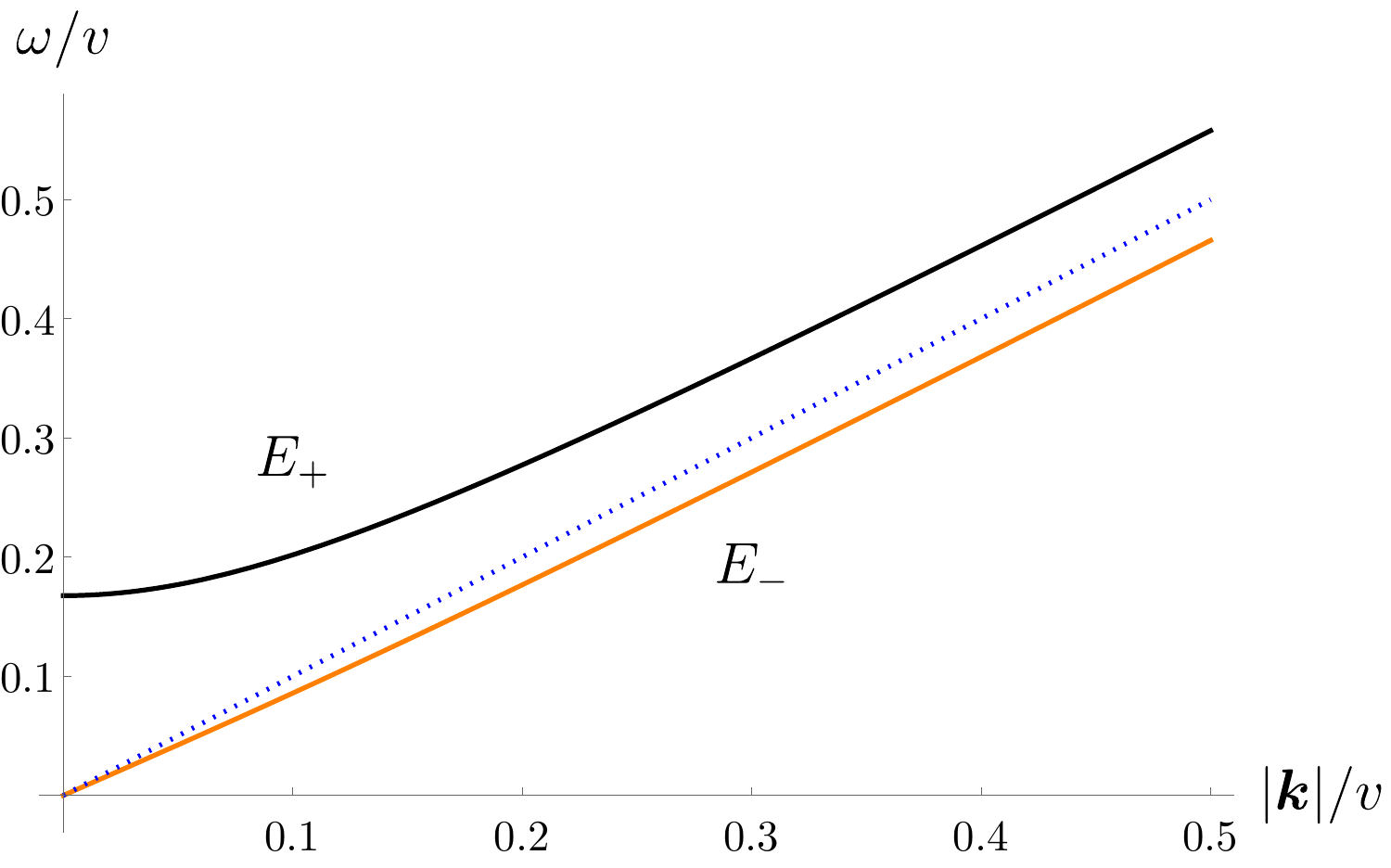}
    \caption{Plot of the dispersion relations of the $+$ (black) and $-$ (orange) states in units of $v$ for $\lambda=10^{-2}$ and $\mu/v = 3/10$. The blue dotted line is the null curve $\omega(\boldsymbol{k}) = |\boldsymbol{k}|$, and we have $E_\pm(\boldsymbol{k}) = |\boldsymbol{k}| \pm \mu^2/2v + \mathcal{O}\left( |\boldsymbol{k}|^{-1} \right)$ as $|\boldsymbol{k}| \rightarrow \infty$. Thus the $+$ branch is always timelike while the $-$ branch is spacelike.}
    \label{fig:Dispersion}
\end{figure}

We start with the renormalizable Lagrangian (we use the mostly-minus signature)
\begin{equation} \label{eq:1}
    \mathcal L = \partial \Phi^{\dagger} \cdot \partial \Phi + m^2 \, \Phi^{\dagger} \Phi - \lambda (\Phi^{\dagger} \Phi)^2 \,.
\end{equation} 
We set
\begin{equation}\label{eq:2}
    \Phi = \frac{\rho}{\sqrt{2}} e^{i \theta /v} \,,
\end{equation} 
where $v$ is a constant with dimensions of energy that will be defined in \eqref{eq:6}. This way we obtain 
\begin{equation}\label{eq:3}
    \mathcal L = \frac{1}{2} (\partial \rho)^2 + \frac{1}{2 v^{2}} (\partial \theta)^{2} \rho^{2} + \frac{m^2}{2} \rho^2 - \frac{\lambda}{4} \rho^4 \,.
\end{equation}   
We are going to study this model around a time-dependent solution $\theta = \mu^2 t/2$, i.e.~with spontaneous breaking of Lorentz invariance. Expanding as $\theta = \mu^2 t/2 + \pi$, we get
\begin{equation}\label{eq:4}
    \begin{aligned}
    \mathcal L &= \frac{1}{2} (\partial \rho)^{2} + \frac{1}{2v^2} \left(\mu^{2} \dot \pi + (\partial \pi)^2 \right) \rho^{2} + \frac{1}{2} \left( m^2 + \frac{\mu^4}{4 v^2} \right) \rho^2 - \frac{\lambda}{4} \rho^4 \,,
\end{aligned} 
\end{equation}
which can be rewritten as 
\begin{equation}\label{eq:5}
    \mathcal L = \frac{1}{2} (\partial \rho)^{2} + \frac{1}{2v^2} \left(\mu^{2} \dot \pi + (\partial \pi)^2 \right) \rho^{2} - \frac{\lambda}{4} (\rho^{2} - v^{2})^{2} + \text{constant}\,,
\end{equation}   
where $v$ is implicitly defined through 
\begin{equation}\label{eq:6}
    v^2 = \frac{1}{\lambda} \left( m^2 + \frac{\mu^4}{4 v^2} \right) \,.
\end{equation}
It is useful to notice that for $\mu \ll  m/\lambda^{1/4}$, $v \approx m/\lambda^{1/2}$, i.e.~the Lorentz invariant result, while for $\mu \gg m/\lambda^{1/4}$, $v \approx \mu/(4 \lambda)^{1/4}$.
Expanding around the minimum by setting $\rho = v + h$, up to the constant and a total derivative we find
\begin{equation}\label{eq:Lagrangianexpanded}
    \mathcal L = \frac{1}{2} (\partial h)^{2} + \frac{1}{2} (\partial \pi)^{2} + \frac{1}{2v^2} \left(\mu^{2} \dot \pi + (\partial \pi)^2 \right) (h^2 + 2vh) - \frac{\lambda}{4} (h^2 + 2 v h)^{2} \,.
\end{equation}  
Let us focus on the quadratic part of the Lagrangian,
\begin{equation}\label{eq:quadraticL}
    \mathcal L_{(2)} = \frac{1}{2} (\partial h)^{2} + \frac{1}{2} (\partial \pi)^{2} + \frac{\mu^2}{v} \dot{\pi} h - \lambda v^2 h^2 \,,
\end{equation} 
which in Fourier space reads
\begin{equation}\label{eq:EOMmatrix}
    \begin{aligned}
     \mathcal L_{(2)} & = \frac{1}{2} \left ( \begin{array}{c c}
     \tilde \pi_{-k}   & \tilde h_{-k}
\end{array} \right  ) \left( \begin{array}{cc}
    k^2  &  i \mu^2 \omega/v \\
    -i \mu^2 \omega/v  & k^2 -M^2
\end{array} \right) \left ( \begin{array}{c}
      \tilde \pi_k\\
      \tilde h_k  
\end{array} \right ) \,,
\end{aligned}  
\end{equation} 
where
\begin{equation}
    M^2 \equiv 2 \lambda v^2 \,.
\end{equation}
By looking at the zeros of the determinant one can find the spectrum of the theory, namely one can find the dispersion relations for the various particles:
\begin{equation}\label{eq:poldispersion}
     (\omega^2 - \boldsymbol{k}^2)(\omega^2 - \boldsymbol{k}^2 - M^2) -  \frac{\mu^4 \omega^2}{ v^2} = 0\,.
\end{equation}
There are two branches of solutions,
\begin{equation} \label{eq:disprelation}
    \omega^2 = E_\pm (\boldsymbol{k})^2 \equiv \boldsymbol{k}^2 + \frac{1}{2} \left( M^2 + \frac{\mu^4}{v^2} \right) \pm \sqrt{ \frac{\mu^4}{v^2} \boldsymbol{k}^2 + \frac{1}{4} \left( M^2 + \frac{\mu^4}{v^2} \right)^2 } \,.
\end{equation}
Thus the Fock space contains two types of particles which are labelled by their three-momentum $\boldsymbol{k}$ and by an extra label $\pm$ indicating which dispersion relates their energy and momentum, see figure \ref{fig:Dispersion}.  The $-$ state is the Goldstone: for small enough $\boldsymbol{k}$ its dispersion relation is linear, with a speed of propagation smaller than unity. The $+$ state is gapped and, for energies much smaller than this gap, one integrates out this state and writes an EFT for the Goldstone only: we will come back to this in section \ref{sec:lowenergies}.  

Now that we have established what the asymptotic states are we must choose which fields to use to interpolate such states. One possibility is to use $h$ and $\pi$. They clearly can interpolate both $\pm$ states. The purpose of the next section is to quantify how much they actually do.

\subsection{Canonical quantization} \label{seccanonical-quantisation}
We begin by introducing the ladder operators of the asymptotic states with commutation relations
\begin{equation}\label{eq:14}
    [a_l(\boldsymbol{k}),a^{\dagger}_m(\boldsymbol{q})]=(2\pi)^3 2 E_l(\boldsymbol{k})\delta_{lm} \delta^{(3)}(\boldsymbol{k}-\boldsymbol{q})
\end{equation}
so that a single particle state of type $l$ with three-momentum $\boldsymbol{k}$ is obtained by
\begin{equation}
    \ket{l,\boldsymbol{k}} = a^{\dagger}_l(\boldsymbol{k})\ket{0} \,, ~~ l \in \{-,+\} \,.
\end{equation}
The fields can be written as\footnote{$\phi^\pi \equiv \pi, \phi^h \equiv h$.}
\begin{equation}\label{eq:Phia}
    \phi^a(t,\boldsymbol{x}) \equiv \sum_{l = \pm} \int \frac{\di^3 \boldsymbol{k}}{(2\pi)^3 2 E_l(\boldsymbol{k})}\left( Z^a_l(\boldsymbol{k}) a_l(\boldsymbol{k})e^{-i (E_l(\boldsymbol{k})t-\boldsymbol{k}\cdot \boldsymbol{x})}+ \text{h.c.} \right) \,, ~~ a \in \{\pi, h\} \,.
\end{equation}
We will now obtain an expression for the $Z$-factors using the canonical commutation relations and the free equations of motion.
The momenta conjugate to $\pi$ and $h$ are
\begin{equation}\label{eq:15}
    \Pi^{\pi}=\dot{\pi}+\frac{\mu^2}{2v}h \,, ~~ \Pi^{h}=\dot{h}-\frac{\mu^2}{2v}\pi \,,
\end{equation}
so that\footnote{$\epsilon^{\pi h} \equiv 1 \equiv -\epsilon^{h\pi }$ and zero otherwise.}
\begin{equation}
    \Pi^b(t,\boldsymbol{x}) = \sum_{l = \pm} \int \frac{\di^3 \boldsymbol{k}}{(2\pi)^3 2 E_l(\boldsymbol{k})}\left[ \left( - i E_l(\boldsymbol{k}) Z^b_l(\boldsymbol{k}) + \epsilon^{bc} \frac{\mu^2}{2v} Z^c_l(\boldsymbol{k}) \right) a_l(\boldsymbol{k})e^{-i (E_l(\boldsymbol{k})t-\boldsymbol{k}\cdot \boldsymbol{x})}+ \text{h.c.} \right] \,.
\end{equation}
The equal-time commutation relations are
\begin{equation}\label{eq:17}
\begin{aligned}
&[\phi^a(t,\boldsymbol{x}),\phi^b(t,\boldsymbol{y})]=0 \,, \\
&[\phi^a(t,\boldsymbol{x}),\Pi^b(t,\boldsymbol{y})]=i\delta^{ab}\delta^{(3)}(\boldsymbol{x}-\boldsymbol{y}) \,, \\&[\Pi^a(t,\boldsymbol{x}),\Pi^b(t,\boldsymbol{y})]=0 \,.
    \end{aligned}
\end{equation}
The first and second equations read\footnote{Here we use invariance under spatial rotations -- that $Z^a_l(\boldsymbol{k})$ is only a function of $|\boldsymbol{k}|$.}
\begin{equation}\label{eq:CCR}
    \begin{aligned}
        &\sum_{l = \pm}\frac{Z^a_l \bar{Z}^b_l - \bar{Z}^a_lZ^b_l}{E_l} = 0 \,, \\
        &\sum_{l = \pm}Z^a_l \bar{Z}^b_l + \bar{Z}^a_l Z^b_l =  2\delta^{ab} \,,
        \end{aligned}
\end{equation}
where we have suppressed the $\boldsymbol{k}$-dependence and indicated with $\bar Z$ the complex conjugate of $Z$. The third equation boils down to
\begin{equation} \label{eq20}
   \sum_{l = \pm} \frac{E_l}{2} \left( Z^a_l \bar{Z}^b_l - \bar{Z}^a_l Z^b_l \right) = -i \frac{\mu^2}{v}\epsilon^{ab} \,.
\end{equation}
The free equations of motion imply that for fixed $l$ the vector $Z^a_l$ is an eigenvector of the matrix in Eq.~\eqref{eq:EOMmatrix}. This gives
\begin{equation}\label{eq:EoMZ}
	Z^h_l = \frac{i E_l \mu^2 }{v k_l^2} Z^\pi_l \,,
\end{equation}
where $k^2_l=E_l(\boldsymbol{k})^2 - \boldsymbol{k}^2$ and this equation holds for $l \in \{-,+\}$ separately. Using Eq.~\eqref{eq:EoMZ} in \eqref{eq:CCR} gives
\begin{equation}
	\sum_{l = \pm} \frac{|Z^\pi_l|^2}{k_l^2 - M^2} = 0 \,, ~~ \sum_{l = \pm} |Z^\pi_l|^2 = 1 \,,
\end{equation}
whose solutions are
\begin{equation}
	|Z^\pi_l|^2 =l \frac{k_l^2-M^2 }{k_+^2 - k_-^2} \,, ~~ l \in \{-,+\} \,.
\end{equation}
Explicitly in terms of the spatial momentum,
\begin{align}
	|Z^{\pi}_{\pm}(\boldsymbol{k})|^2 &= \frac{1}{2} \pm \frac{\mu^4/v^2 - M^2}{2\sqrt{4\mu^4 \boldsymbol{k}^2/v^2 + \left( \mu^4/v^2 + M^2 \right)^2}} \,, \\
	|Z^h_\pm(\boldsymbol{k})|^2 &= \frac{1}{2} \pm \frac{\mu^4/v^2+M^2}{2\sqrt{4\mu^4 \boldsymbol{k}^2/v^2 + \left(\mu^4/v^2 + M^2 \right)^2}} \,. \label{eq:21}
\end{align}
With the relations \eqref{eq:EoMZ}, there are two remaining unfixed phases in the solution. These correspond to our freedom in \eqref{eq:Phia} to rotate $a_\pm(\boldsymbol{k})$. Choosing $Z^{\pi}_{-}$ and $Z^{h}_{+}$ to be positive real, the explicit expressions for the interpolating factors are
\begin{align}
	Z^{\pi}_{-}(\boldsymbol{k}) &= \sqrt{\frac{1}{2} - \frac{\mu^4/v^2 - M^2}{2\sqrt{4\mu^4 \boldsymbol{k}^2/v^2 + \left( \mu^4/v^2 + M^2 \right)^2}}} \,, \\
	Z^h_-(\boldsymbol{k}) &= -i \sqrt{\frac{1}{2} - \frac{\mu^4/v^2+M^2}{2\sqrt{4\mu^4 \boldsymbol{k}^2/v^2 + \left(\mu^4/v^2 + M^2 \right)^2}}} \,, \\
	Z^{\pi}_{+}(\boldsymbol{k}) &= - i\sqrt{\frac{1}{2} + \frac{\mu^4/v^2 - M^2}{2\sqrt{4\mu^4 \boldsymbol{k}^2/v^2 + \left( \mu^4/v^2 + M^2 \right)^2}}} \,, \\
	Z^h_+(\boldsymbol{k}) &= \sqrt{\frac{1}{2} + \frac{\mu^4/v^2+M^2}{2\sqrt{4\mu^4 \boldsymbol{k}^2/v^2 + \left(\mu^4/v^2 + M^2 \right)^2}}} \,.
	 \label{eq:22}
\end{align}
From these expressions we see
\begin{equation} \label{eq:highE_Z}
   Z^\pi_{-} (\boldsymbol{k}) \sim Z^h_{+} (\boldsymbol{k})  \sim \frac{1}{\sqrt{2}} \,, ~~ Z^\pi_{+} (\boldsymbol{k}) \sim Z^h_{-} (\boldsymbol{k})  \sim - \frac{i}{\sqrt{2}} ~~ \text{as } |\boldsymbol{k}| \rightarrow \infty \,.
\end{equation}
This means that at high energies the two fields $\pi, h$ interpolate the two asymptotic states $-,+$ equally well. This can also be understood by considering the quadratic Lagrangian \eqref{eq:quadraticL} and noticing that, in the high energy limit, the splitting is dominated by the one-derivative operator $(\mu^2/v) h \dot \pi$ while the mass can be neglected. So at high energies the quadratic Lagrangian reads 
\begin{equation}
    \mathcal L_{(2)}  \sim \frac{1}{2} \left ( \begin{array}{c c}
      \pi   &  h
\end{array} \right) \left( \begin{array}{cc}
    -\partial^2  &  -(\mu^2/v) \partial_t \\
    (\mu^2/v) \partial_t  &  -\partial^2
\end{array} \right) \left ( \begin{array}{c}
      \pi\\
       h
\end{array} \right )  
\end{equation}
which can be diagonalized yielding 
\begin{equation}
 \mathcal L_{(2)} \sim \frac{1}{2} \times \frac{1}{\sqrt{2}} \left ( \begin{array}{c c}
      \pi  - i h  &  h - i \pi
\end{array} \right) \left( \begin{array}{cc}
    -\partial^2 + i (\mu^2/v) \partial_t  &  0 \\
    0  & -\partial^2 - i (\mu^2/v) \partial_t
\end{array} \right) \frac{1}{\sqrt{2}} \left (\begin{array}{c}
      \pi + i h \\
       h + i\pi
\end{array} \right ) . 
\end{equation}
In the two eigenvalues we recognize the high energy limits of the dispersion relations \eqref{eq:disprelation} and in the eigenvectors we recognize the high energy limits of the $Z$-factors \eqref{eq:highE_Z}. Since the leading splitting operator $(\mu^2/v) h \dot \pi$ is Lorentz-breaking, going to very high energies one does not recover the same eigenvectors as in the Lorentz invariant limit $\mu \rightarrow 0$. At low energies, conversely, the splitting is dominated by the mass. The crossover between the two regimes happens at
\be \label{omegacEq}
\frac{\mu^2\omega}{v}\sim M^2\quad\Rightarrow \quad \omega_c\sim \frac{M^2 v}{\mu^2} \sim \lambda \frac{v^3}{\mu^2} \,.
\ee
For $\mu\lesssim m/\lambda^{1/4}$, this crossover scale is $\omega_c\sim m^3/(\mu^2\lambda^{1/2})$, otherwise, for $\mu \gtrsim m/\lambda^{1/4}$, $\omega_c \sim \lambda^{1/4}\mu$.

\section{\label{secLSZ}LSZ reduction formula and lack of analyticity}
In this section we will derive the LSZ reduction formula \cite{Lehmann:1954rq}, following Weinberg's ``polology" arguments \cite{Weinberg:1995mt}.\footnote{We acknowledge Riccardo Rattazzi for clarifying certain aspects of Weinberg's ``polology" arguments and their link to the LSZ reduction formula, during the ``Advanced Quantum Field Theory" course given at EPFL in 2020.} This is well-known material in the Lorentz invariant case, and we will highlight the differences in the absence of Lorentz invariance.

We want to prove the following relation between time-ordered Green's functions and $S$-matrix elements:
\be\label{eq:LSZpolology}
\begin{split}
\prod_{i=1}^n \int \di^4 y_i \, e^{i p_i \cdot y_i} \prod_{j=1}^m \int \di^4 x_j \, e^{-i k_j \cdot x_j} \langle 0 | T(\pi(y_1)\ldots \pi(y_n) \pi(x_1) \ldots \pi(x_m))| 0\rangle \cong \\
\prod_{i=1}^n \frac{i Z_-^\pi(\boldsymbol{p}_i)}{(p_i^0)^2 - E_-(\boldsymbol{p}_i)^2 + i \varepsilon} \prod_{j=1}^m \frac{i \bar Z_-^\pi(\boldsymbol{k}_j)}{(k_j^0)^2 - E_-(\boldsymbol{k}_j)^2 + i \varepsilon} \langle \boldsymbol{p}_1 \ldots \boldsymbol{p}_n | S | \boldsymbol{k}_1 \ldots \boldsymbol{k}_m\rangle \,.
\end{split}
\ee
The $\cong$ symbol indicates the equality holds in the limit all particles go on-shell: $p_i^0 \to E_-(\boldsymbol{p}_i)$ and $k_j^0 \to E_-(\boldsymbol{k}_j)$. Notice we are taking the convention that $p_i^0, k_j^0 > 0$. The matrix elements were defined in \eqref{eq:Phia} as
\begin{equation}\label{eq:24}
    Z_\pm^a(\boldsymbol{k}) \equiv \braket{0 | \phi^a(0) | \pm, k } \,, ~~ a \in \{h,\pi\} \,.
\end{equation}
(In the standard Lorentz invariant treatment, these would reduce to a factor $\sqrt{Z}$, which describes the wavefunction normalization.) 
Here we are considering the scattering of $-$ particles; if one were interested in $+$ particles, one should use $Z_+$'s and also use the dispersion relations $E_+(\, \cdot \,)^2$. In this way of deriving the LSZ reduction formula, one is free to use any field to interpolate the particles of interest provided there is a nonzero overlap. For instance, one could write a similar formula by replacing some (or all) the fields $\pi$ with $h$. Since the overlap between a $-$ particle and the $h$-field goes to zero at low energies (see \eqref{eq:22}), it is more natural to use $\pi$.

To prove \eqref{eq:LSZpolology} one focuses on the regions of integration $y_i^0 \to + \infty$ and $x_i^0 \to - \infty$ since this is the region that gives rise to the poles \cite{Weinberg:1995mt}. In this regime all the $y$'s are after all the $x$'s and the time-ordered product factorizes as 
\be\label{eq:25}
\langle 0 | T(\pi(y_1)\ldots \pi(y_n))T( \pi(x_1) \ldots \pi(x_m))| 0\rangle \,.
\ee
We can thus insert a complete set of states (twice) in the middle,
\be\label{eq:26}
\sum_{\alpha,\beta}\langle 0 | T(\pi(y_1)\ldots \pi(y_n)) |\alpha_{\rm out} \rangle \langle\alpha_{\rm out}  |\beta_{\rm in} \rangle \langle\beta_{\rm in}|T( \pi(x_1) \ldots \pi(x_m))| 0\rangle \,.
\ee
The object $\langle\alpha_{\rm out}  |\beta_{\rm in} \rangle $ is the $S$-matrix. In the late-time limit, after the wave packets of the various particles are separated, the matrix element on the left is dominated by a state with $n$ particles:
\be\label{eq:27}
\begin{split}
\lim_{y_i^0 \to + \infty} \langle 0 | T(\pi(y_1)\ldots \pi(y_n)) |\alpha_{\rm out} \rangle  & \cong \langle 0 | T(\pi(y_1)\ldots \pi(y_n)) | \boldsymbol{q}_1 \ldots \boldsymbol{q}_n \rangle_{\rm out} \cong \\ \cong \prod_{i=1}^n  \langle 0 | \pi(y_i)| \boldsymbol{q}_i \rangle_{\rm out}
\cong & \prod_{i=1}^n Z_-^\pi(\boldsymbol{q}_i) e^{-i q_i \cdot y_i} \,.
\end{split}
\ee
(The resolution of identity contains also $+$ particles, but those would not contribute to the pole when one goes on-shell for $-$ states.) Notice that if the theory were invariant under boosts, $Z$ would not depend on $\boldsymbol{q}$: as we will see, this is a crucial difference. (We have been sloppy here with permutations, which will eventually cancel with the normalization of states with identical particles.) Each term in the product gives
\be\label{eq:28}
\int \di^4 y_i \; e^{i p_i \cdot y_i} Z_-^\pi(\boldsymbol{q}_i) e^{-i q_i \cdot y_i} = \int \di y_i^0  (2\pi)^3 \delta^{(3)} (\boldsymbol{p}_i-\boldsymbol{q}_i)Z_-^\pi(\boldsymbol{q}_i) e^{i (p_i^0 - q_i^0) y_i^0}\,,
\ee
where we did the integral over the spatial variables. Then doing the integral over the phase space, $\int {\rm d}^3\boldsymbol{q}_i/((2\pi)^3 2 E_-(\boldsymbol{q}_i))$, one gets
\be\label{eq:29}
 \int \di y_i^0 \frac{1}{2 E_-(\boldsymbol{p}_i)}  Z_-^\pi(\boldsymbol{p}_i) e^{i (p_i^0 - E_-(\boldsymbol{p}_i)) y_i^0} \cong \frac{i Z_-^\pi(\boldsymbol{p}_i)}{(p_i^0)^2 - E_-(\boldsymbol{p}_i)^2 + i \varepsilon} \,,
\ee
where the last equality is valid up to terms that vanish as one goes close to the pole. To make the time integral well-defined one has to slightly deform the contour of integration, hence the $i \varepsilon$. The same arguments work for the incoming particles, proving the LSZ formula \eqref{eq:LSZpolology}.

We can now proceed to study the analyticity of the $2 \to 2$ $S$-matrix. In this case the LSZ formula, in which only one in-going and one out-going leg is reduced, gives
\be\label{eq:30}
S = - \int \di^4 x \, \di^4 y \,e^{i (q_2 \cdot y - q_1 \cdot x)} \frac{-\partial_{y^0}^2 - E_-(-i \partial_{y_i})^2}{Z_-^\pi(-i \partial_{y_i})} \frac{-\partial_{x^0}^2 - E_-(-i \partial_{x_i})^2}{\bar Z_-^\pi(-i \partial_{x_i})} \langle \boldsymbol{p}_2 | T(\pi(y) \pi(x)) |\boldsymbol{p}_1\rangle \,.
\ee
We slightly changed the notation calling $\boldsymbol{q}_1$ and $\boldsymbol{p}_1$ the incoming momenta ($\boldsymbol{q}_2$ and $\boldsymbol{p}_2$ are outgoing). This simplifies the comparison with Itzykson \& Zuber's book \cite{Itzykson:1980rh}, whose logic we will follow. As in the Lorentz invariant case one can replace the time-ordered correlation function with the retarded two-point function: the difference only contributes to a disconnected piece of the $S$-matrix. Therefore one has
\be\label{eq:31}
S = - \int \di^4 x \, \di^4 y \,e^{i (q_2 \cdot y - q_1 \cdot x)} \frac{-\partial_{y^0}^2 - E_-(-i \partial_{y_i})^2}{Z_-^\pi(-i \partial_{y_i})} \frac{-\partial_{x^0}^2 - E_-(-i  \partial_{x_i})^2}{\bar Z_-^\pi(-i \partial_{x_i})} \langle \boldsymbol{p}_2 | \theta(y^0-x^0)[\pi(y), \pi(x)] |\boldsymbol{p}_1\rangle \,.
\ee
Using translation invariance one can then factor out the delta function of energy and momentum conservation,
\be\label{eq:32}
S = (2\pi)^4 \delta^{(4)} (p_2 +q_2 -p_1-q_1) i \mathcal{M} \,,
\ee
so that, in terms of $q \equiv \frac12 (q_1 + q_2)$, one has
\be\label{eq:Tdefinition}
\mathcal{M} = i \int \di^4 z\, e^{i q \cdot z}   \langle \boldsymbol{p}_2 | \theta(z^0) \left[\frac{\partial_{z^0/2}^2 + E_-(-i \partial_{z_i/2})^2}{Z_-^\pi(-i \partial_{z_i/2})}\pi \left( \frac{z}{2} \right), \frac{\partial_{z^0/2}^2 + E_-(-i \partial_{z_i/2})^2}{\bar Z_-^\pi(-i \partial_{z_i/2})}\pi \left( -\frac{z}{2} \right) \right] |\boldsymbol{p}_1\rangle \,.
\ee
In the Lorentz invariant case the $Z$'s are just constants and the differential operators are the Klein-Gordon ones, $\Box +m^2$. Since these operators are local, the integrand is nonzero only for $z^\mu$ in the forward lightcone. Therefore $\mathcal{M}(q^\mu)$ is analytic if $\mathrm{Im}\;q^\mu$ lies in the forward lightcone. (In this region all the derivatives of $\mathcal{M}$ are convergent integrals, assuming polynomial boundedness of the matrix elements, and it thus defines an analytic function.) This primitive domain of analyticity is the stepping stone for the use of dispersive arguments in the $S$-matrix \cite{Sommer:1970mr,Itzykson:1980rh}. Here, however, the differential operators acting on the fields are non-local. In particular, from the explicit expressions of $Z_-^\pi$ and $E_-^2$ one sees the presence of branch points for complex spatial momenta. These singularities are present for any value of the time components of the momenta. This means there is no cone of analyticity for $\mathrm{Im}\;q^\mu$, contrary to the Lorentz invariant case. The only guaranteed region of analyticity, which is universal and independent of the model at hand, is for $\mathrm{Im}\;q^0 >0$: this is just a consequence of the fact the integrand above is retarded, i.e.~different from zero only for $z^0 \ge 0$.

Another way to see the absence of analyticity is to consider fields built only out of creation/annihilation operators of $-$ (or respectively $+$) states:
\be
\varphi_l(t,\boldsymbol{x})\equiv \int \frac{\di^3 \boldsymbol{k}}{(2\pi)^3 2 E_l(\boldsymbol{k})}\left( a_l(\boldsymbol{k})e^{-i (E_l(\boldsymbol{k})t-\boldsymbol{k}\cdot \boldsymbol{x})}+ \text{h.c.} \right) \,, ~ l \in \{-,+\} \,.
\ee
These fields interpolate only one of the two eigenstates, so that $\langle 0 | \varphi_l(0)| l,q \rangle_{\rm out} = 1$. The $\varphi$'s however are not microcausal, i.e.~{\em they do not commute outside the lightcone}. The simplest way to see this is to look at the Fourier representation of the retarded Green's function,
\be
G_{R,l}(\omega,\boldsymbol{k})= \frac{i}{(\omega+ i \varepsilon)^2 - E_l(\boldsymbol{k})^2} \,.
\ee
A necessary and sufficient condition for the vanishing of the retarded Green's function outside the forward lightcone is that its Fourier representation is analytic when the imaginary part of the 4-vector $(\omega,\boldsymbol{k})$ is in the forward lightcone. This property does not hold here since $E_l(\boldsymbol{k})$ has branch points, as we saw above. On the other hand, the fields $\pi(x)$ and $h(x)$ commute outside the lightcone.\footnote{Notice that $\pi(x)$ is not a Lorentz scalar, since Lorentz invariance is spontaneously broken. However the relation between the original Lorentz invariant field $\Phi(x)$ and $\pi(x), h(x)$ is local, so that microcausality is preserved.} In order to create a single energy eigenstate, one has to act non-locally in the fields and this is the ultimate reason for the absence of analyticity.

The relation between the $S$-matrix and correlation functions is uniquely defined only in the limit when the external momenta are real and on-shell. Therefore the definition of Eq.~\eqref{eq:Tdefinition} is ambiguous for generic complex $q^\mu$.\footnote{For fixed $\boldsymbol{p}_1$ and $\boldsymbol{p}_2$ the $S$-matrix depends on 3 real variables, for instance the vector $\boldsymbol{q}_1$ (in the presence of Lorentz invariance it would depend on a single real variable). Therefore there is no unique analytic extension to a function of $q^\mu \in \mathbb{C}^4$.} In appendix \ref{app2LSZ} we derive an LSZ reduction formula starting from the explicit expression of the creation and annihilation operators of the $+$ and $-$ states. The final reduction formula gives the same $S$-matrix, but it is different from the one obtained above. The conclusion about the lack of analyticity remains the same, however. Given this ambiguity, one may wonder whether a ``clever" LSZ procedure exists that gives an analytic off-shell $S$-matrix. In particular analyticity is preserved if one acts on each external leg with a local differential operator, i.e.~a polynomial in Fourier space. This is indeed what happens in the Lorentz invariant case. We will now prove that this does not work: there is no way to extract the correct $S$-matrix by acting with local differential operators. 

Let us assume that such an operator exists. This corresponds to a polynomial $P(\omega, |\boldsymbol{k}|^2)$ that multiplies each external leg. (For rotational invariance $P$ must depend only on $|\boldsymbol{k}|^2$.) In order to give the correct $S$-matrix this polynomial must satisfy
\be
P(\omega, |\boldsymbol{k}|^2) = \frac{2 E_-(|\boldsymbol{k}|)}{Z_-^\pi(|\boldsymbol{k}|)} \left( \omega - E_-(|\boldsymbol{k}|) \right) + \mathcal{O} \left( \left( \omega - E_-(|\boldsymbol{k}|) \right)^2 \right) ~~ \text{as } \omega \rightarrow E_-(|\boldsymbol{k}|) \,.
\ee
At fixed $|\boldsymbol{k}|$, the polynomial in $\omega$ must have $E_-(|\boldsymbol{k}|)$ as a simple root. This is surely possible: indeed the dispersion relations in the model at hand are solutions of the polynomial equation \eqref{eq:poldispersion}. To get also the correct normalization, $P$ must satisfy
\be
\left.\frac{\partial}{\partial\omega} P(\omega, |\boldsymbol{k}|^2) \right|_{\omega = E_-(|\boldsymbol{k}|)} = \frac{2 E_-(|\boldsymbol{k}|)}{Z_-^\pi(|\boldsymbol{k}|)} \,.
\ee
Taking the partial derivative with respect to $\omega$ of the polynomial $P$ gives another polynomial $Q(\omega,|\boldsymbol{k}|^2)$, which should satisfy
\be
Q(\omega = E_-(|\boldsymbol{k}|), |\boldsymbol{k}|^2) = \frac{2 E_-(|\boldsymbol{k}|)}{Z_-^\pi(|\boldsymbol{k}|)} \,.
\ee
Let us look at the two sides of this equation as a function of the complex variable $|\boldsymbol{k}|^2$. The RHS is analytic around the positive real axis $|\boldsymbol{k}|^2 > 0$: if the two sides of the equation above coincide for physical values $|\boldsymbol{k}|^2 > 0$, they must have the same analytic extension in the complex $|\boldsymbol{k}|^2$ plane (properly extended to take into account branch points). However it is impossible that the two functions coincide in the complex plane. The RHS has poles at the zeros of $Z_-^\pi(|\boldsymbol{k}|)$, while the LHS has no poles being a polynomial in $|\boldsymbol{k}|^2$ and $E_-(|\boldsymbol{k}|)$ (see its explicit expression Eq.~\eqref{eq:disprelation}). Therefore $P$ cannot exist.

In conclusion, one cannot extract the $S$-matrix from correlation functions acting with local differential operators, preserving in this way the analyticity of the retarded Green's function. One cannot exclude some more exotic way of defining the $S$-matrix for complex momenta that preserves analyticity.

\section{\label{secSmatrix}The \texorpdfstring{$S$}{S}-matrix}
The object of interest in this work is the scattering amplitude for the process $\ket{-,\boldsymbol{p}_1} \ket{-,\boldsymbol{q}_1}$ into $\ket{-,\boldsymbol{p}_2} \ket{-,\boldsymbol{q}_2}$. In what follows we will denote this scattering amplitude by $\mathcal M$ (details about the calculation are deferred to appendix \ref{appFeynman}.)
\begin{figure}[ht]
    \centering
    \includegraphics[width=0.43\textwidth]{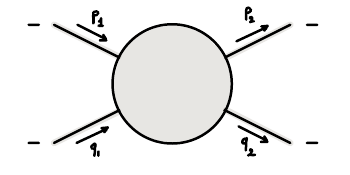}
\end{figure}
This scattering amplitude has four contributions, from contact interactions and from exchange diagrams in the three channels: 
\begin{equation}\label{eq:SMatrix}
    \mathcal M = \mathcal M_c + \mathcal M_s + \mathcal M_t + \mathcal M_u \,.
\end{equation}
These terms read 
\begin{equation}\label{eq:35}
    \begin{aligned}
   \mathcal{M}_c &= -  \frac{\mu^4}{v^4}  \prod_{i=1}^2 \left (  \frac{ Z^\pi_{-} (\boldsymbol{p}_i)^{-1}  }{(p^{0}_i)^2 - E_+(\boldsymbol{p}_i)^2} \frac{ Z^\pi_{-} (\boldsymbol{q}_i)^{-1}  }{(q^{0}_i)^2 - E_+(\boldsymbol{q}_i)^2}  \right )   \Big [6 \lambda \mu^4 p_1^0 q_1^0 p_2^0 q_2^0 +\\
   &- 2 p_1\cdot q_1 (p_1^2 - M^2) (q_1^2 - M^2) p_2^0 q_2^0- 2 p_1\cdot q_2 (p_1^2 - M^2) (q_2^2 - M^2) q_1^0 p_2^0 +
   \\
   & - 2 p_1\cdot p_2 (p_1^2 - M^2) (p_2^2 - M^2) q_1^0 q_2^0 - 2 q_1\cdot q_2 (q_1^2 - M^2) (q_2^2 - M^2) p_1^0 p_2^0 + \\ & - 2 p_2\cdot q_1 (p_2^2 - M^2) (q_1^2 - M^2) p_1^0 q_2^0 - 2 q_2\cdot p_2 (p_2^2 - M^2) (q_2^2 - M^2) p_1^0 q_1^0 \Big] \,,
\end{aligned}  
\end{equation}
\begin{equation}\label{eq:36}
    \begin{aligned}
    & \mathcal M_s  =  \prod_{i=1}^2 \left (  \frac{ Z^\pi_{-} (\boldsymbol{p}_i)^{-1}  }{(p^{0}_i)^2 - E_+(\boldsymbol{p}_i)^2} \frac{ Z^\pi_{-} (\boldsymbol{q}_i)^{-1}  }{(q^{0}_i)^2 - E_+(\boldsymbol{q}_i)^2}  \right ) \frac{ v^{-9}}{(p_1 + q_1)^2 ( (p_1+q_1)^2 - M^2 )  - \frac{\mu^4}{v^2} (p_1^0+q_1^0)^2} \\ 
    &  \Big [2 \lambda v^3 \mu^4  B(p_1,q_1) B(p_2,q_2) + [B(p_1,q_1) A(p_2,q_2)  +B(p_2,q_2) A(p_1,q_1)] \mu^4 (p_1^0+q_1^0)- \\ & v [A(p_1,q_1) A(p_2,q_2)  + \mu^4 B(p_1,q_1) B(p_2,q_2) ] (p_1 + q_1)^2  \Big] \,,
\end{aligned}
\end{equation} 
\begin{equation}\label{eq:37}
    \begin{aligned}
    & \mathcal M_t  = \prod_{i=1}^2 \left (  \frac{ Z^\pi_{-} (\boldsymbol{p}_i)^{-1}  }{(p^{0}_i)^2 - E_+(\boldsymbol{p}_i)^2} \frac{ Z^\pi_{-} (\boldsymbol{q}_i)^{-1}  }{(q^{0}_i)^2 - E_+(\boldsymbol{q}_i)^2}  \right )  \frac{ v^{-9}}{(q_1 - q_2)^2 ( (q_1-q_2)^2 - M^2 ) - \frac{\mu^4}{v^2} (q_1^0 - q_2^0)^2} \\ 
    &  \Big [2 \lambda v^3 \mu^4  B(q_1,- q_2) B(-p_1,p_2) + [B(q_1,-q_2) A(-p_1,p_2)  +B(-p_1,p_2) A(q_1,-q_2)] \mu^4 (q_1^0 - q_2^0) - \\ & v [A(q_1,-q_2) A(-p_1,p_2)  + \mu^4 B(q_1,-q_2) B(-p_1,p_2) ] (q_1-q_2)^2   \Big] \,,
\end{aligned} 
\end{equation} 
and 
\begin{equation}\label{eq:38}
    \begin{aligned}
   & \mathcal M_u  = \prod_{i=1}^2 \left (  \frac{ Z^\pi_{-} (\boldsymbol{p}_i)^{-1}  }{(p^{0}_i)^2 - E_+(\boldsymbol{p}_i)^2} \frac{ Z^\pi_{-} (\boldsymbol{q}_i)^{-1}  }{(q^{0}_i)^2 - E_+(\boldsymbol{q}_i)^2}  \right ) \frac{ v^{-9}}{(q_1 - p_2)^2 ( (q_1-p_2)^2 - M^2 ) - \frac{\mu^4}{v^2} (q_1^0-p_2^0)^2} \\ 
    & \Big [2 \lambda v^3 \mu^4  B(q_1,-p_2) B(q_2,-p_1) + [B(q_1,-p_2) A(q_2,-p_1)  +B(q_2,-p_1) A(q_1,-p_2)] \mu^4 (q_1^0-p_2^0) - \\ & v [A(q_1,-p_2) A(q_2,-p_1)  + \mu^4 B(q_1,-p_2) B(q_2,-p_1) ] (q_1-p_2)^2   \Big ] \,.
\end{aligned}
\end{equation}
Here, as specified in section~\ref{seccanonical-quantisation}, we have chosen $Z^\pi_-$ to be real, so we do not make distinction between $Z^\pi_-$ and $\bar Z^\pi_-$. In order to write the blocks in a slightly more compact form we have defined the functions 
\begin{equation}\label{eq:39}
    A(p_1,q_1) \equiv 2 v^3 (p_1^2 - M^2) (q_1^2 - M^2) p_1 \cdot q_1 - v \mu^4 p_1^0q_1^0 (p_1^2 + q_1^2 + M^2) \,,
\end{equation}  
and 
\begin{equation}\label{eq:40}
    B(p_1,q_1) \equiv \mu^4 (p_1^0)^2 q_1^0  - 2 v^2 q_1^0 (p_1^2-M^2) (p_1^2 + p_1 \cdot q_1) + (p_1 \leftrightarrow q_1) \,.
\end{equation} 
It is straightforward to check that the amplitude would be analytic in $q \equiv \frac12 (q_1 + q_2)$ for $\mathrm{Im}\;q^\mu$ in the forward lightcone, if the LSZ procedure did not introduce the non-analyticities discussed the previous section. Notice that under the exchange $p_1 \leftrightarrow - q_2$ we have $  \mathcal M_s \leftrightarrow \mathcal  M_t$ and under $p_1 \leftrightarrow - p_2$ we have $  \mathcal M_s \leftrightarrow \mathcal  M_t$, whereas $\mathcal M_c$ is always unchanged. This shows that $\mathcal M$ is invariant under the exchange of the external momenta, i.e. it is crossing symmetric.

\subsection{Interesting limits}
Given the complexity of the scattering amplitude it is useful to study it in some interesting limits.

\subsubsection{\texorpdfstring{$\mu= 0$}{muis0}}
As a sanity check we want to be sure to recover the usual Lorentz invariant form of the $S$-matrix when we switch off the breaking $\mu$. It is straightforward to see that for $\mu =0$,
    \begin{equation}\label{eq:41}
    \begin{aligned}
        & Z_{-}^{\pi} = 1\,, \quad E_{-}(\boldsymbol{k})^2 = \boldsymbol{k}^2\,, \quad  E_{+}(\boldsymbol{k})^2 = \boldsymbol{k}^2 + M^2\,,  \\&A(p_1, q_1)= 2 v^3 M^4 p_1\cdot q_1 \,,\quad B(p_1,q_1)= 2 v^2M^2(p_1^0+q_1^0)p_1\cdot q_1 \,. 
    \end{aligned}
\end{equation}
As a consequence, the various contributions to the tree-level amplitude reduce to
\begin{equation}\label{eq:42}
    \mathcal{M}_c = 0 \,, ~ \mathcal{M}_s=-\frac{s^2}{v^2(s-M^2)} \,, ~ \mathcal{M}_t=-\frac{t^2}{v^2(t-M^2)} \,, ~ \mathcal{M}_u=-\frac{u^2}{v^2(u-M^2)} \,,
\end{equation}
using the usual Maldelstam variables. The full amplitude is thus
\begin{equation}\label{eq:LIAmplitude}
    \mathcal{M}= -\frac{1}{v^2}\left[ \frac{s^2}{s-M^2}+ \frac{t^2}{t-M^2}+\frac{u^2}{u-M^2}\right].
\end{equation}
The amplitude reproduces its Lorentz invariant counterpart as soon as the breaking is switched off.

\subsubsection{High energies}
Another interesting limit to explore is the ``high-energy limit", intended as the limit achieved when the energies of the scattered states are much larger than all the other scales involved in the problem. In order to take this limit, for convenience, we specialize to the kinematical configuration in which the energies of the scattering states are all the same
\begin{equation}\label{eq:44}
    q_1^0 = q_2^0 = p_1^0 = p_2^0 \equiv \omega \,,
\end{equation}
and in which the angle between $\vec{q}_1$ and $\vec{p}_1$ is not zero. The high-energy limit is therefore realized by
\begin{equation}\label{eq:45}
    \lim_{\omega/\Omega \to \infty}\mathcal{M}
\end{equation}
where 
\begin{equation}\label{eq:46}
    \Omega \gg v,M,\mu,\omega_c
\end{equation}
is a reference scale whose value is much larger than all the other scales of the problem. In this limit, the behavior of the different contributions to the amplitude is
\begin{equation}\label{eq:47}
    \begin{aligned}
    \mathcal{M}_c=    &  -\frac{\omega ^2 (\hat{p_1}\cdot \hat{q_1}+\hat{p_1}\cdot \hat{q_2}+\hat{q_1}\cdot \hat{q_2}-3)}{v^2}-\frac{3 \mu ^2 \omega }{v^3} \\
        &+\frac{9 \mu ^8-3 M^4 v^4+(\hat{p_1}\cdot \hat{q_1}+\hat{p_1}\cdot \hat{q_2}+\hat{q_1}\cdot \hat{q_2} )\left(\mu ^4+M^2 v^2\right)^2-6 \lambda  \mu ^4 v^4}{4 \mu ^4 v^4}+  \mathcal{O} \left(\frac{\Omega}{\omega} \right) \,,
    \end{aligned}
\end{equation}
\begin{equation}\label{eq:50}
    \begin{aligned}
       \mathcal{M}_s  =  & \frac{5 \omega ^2 (\hat{p_1}\cdot \hat{q_1}-1)}{2 v^2}+\frac{\omega  \left(6 \mu ^4+M^2 v^2 \hat{p_1}\cdot \hat{q_1}-M^2 v^2\right)}{2 \mu ^2 v^3} \\
    &+\frac{1}{8} \left(16 \lambda +\frac{3 M^4}{\mu ^4}+\frac{8 M^2}{v^2}+\hat{p_1}\cdot \hat{q_1} \left(-\frac{3 M^4}{\mu ^4}-\frac{10 M^2}{v^2}-\frac{5 \mu ^4}{v^4}\right)-\frac{19 \mu ^4}{v^4}\right)+ \mathcal{O} \left(\frac{\Omega}{\omega}\right) \,,
    \end{aligned}
\end{equation}
\begin{equation}\label{eq:49}
    \begin{aligned}
      \mathcal{M}_t=   & \frac{\omega ^2(1-\hat{q_1}\cdot \hat{q_2})}{2 v^2}+\frac{M^2 \omega(1 -  \hat{q_1}\cdot \hat{q_2})}{2 \mu ^2 v} \\
         &+\frac{-\mu ^8+M^4 v^4-16 \mu ^4 M^2 v^2+\hat{q_1}\cdot \hat{q_2}\left(\mu ^8-M^4 v^4+2 \mu ^4 M^2 v^2\right)}{8 \mu ^4 v^4}+ \mathcal{O} \left(\frac{\Omega}{\omega}\right) \,, \\
    \end{aligned}
\end{equation}
\begin{equation}\label{eq:48}
    \begin{aligned}
    \mathcal{M}_u=  &     \frac{\omega ^2(1- \hat{p_1}\cdot \hat{q_2})}{2 v^2}+\frac{M^2 \omega(1 -  \hat{p_1}\cdot \hat{q_2})}{2 \mu ^2 v} \\
         &+\frac{-\mu ^8+M^4 v^4-16 \mu ^4 M^2 v^2+\hat{p_1}\cdot \hat{q_2} \left(\mu ^8-M^4 v^4+2 \mu ^4 M^2 v^2\right)}{8 \mu ^4 v^4}+ \mathcal{O} \left(\frac{\Omega}{\omega}\right) \text{ as } \frac{\omega}{\Omega} \rightarrow \infty \,. \\
    \end{aligned}
\end{equation}
where $\hat{p_i}\cdot \hat{q_j} = \boldsymbol{p}_i \cdot \boldsymbol{q}_j / |\boldsymbol{p}_i||\boldsymbol{q}_j|$.  
Momentum conservation enforces the condition 
\[
1+\hat{p_1}\cdot \hat{q_1}-\hat{q_2}\cdot \hat{q_1}-\hat{p_1}\cdot \hat{q_2}=0 \,,
\]
which ensures the cancellation of the divergent terms in the above blocks, yielding a constant asymptotic value for the full amplitude:
\begin{equation}\label{highenergylimit}
    \mathcal{M}=-2\frac{M^2}{v^2}+ \mathcal{O} \left(\frac{\Omega}{\omega}\right)=-4 \lambda+ \mathcal{O} \left(\frac{\Omega}{\omega}\right) ~~ \text{as } \>\frac{\omega}{\Omega} \rightarrow \infty \,.
\end{equation}
Interestingly, the amplitude asymptotes to the Lorentz-invariant amplitude for the scattering of $U(1)$ eigenstates (quanta of $\Phi$)  in the unbroken phase and not to the $s\to +\infty$ limit of the Lorentz-invariant amplitude that describes the finite-angle scattering of Goldstones in the broken phase\footnote{We thank the authors of \cite{Hui:2023pxc} for pointing out a mistake in the first version of our paper.}. Indeed, making use of \eqref{eq:1}, it is possible to check that the $\Phi \Phi \to \Phi \Phi$ scattering amplitude is given by
\begin{equation}
\mathcal{M}_{\Phi \Phi \to \Phi \Phi}=-4\lambda \;,
\end{equation}
in agreement with \eqref{highenergylimit}.  (On the other hand, the Lorentz-invariant amplitude of the Goldstones, Eq.~\eqref{eq:LIAmplitude}, asymptotes to $-6\lambda$ in the $s \to +\infty$ limit at finite angle.) As discussed at the end of section~\ref{seccanonical-quantisation}, for $\omega \gg \omega_c$ the quanta of $\Phi$ are indeed the correct eigenstates. We refer the reader to reference \cite{Hui:2023pxc} for a more comprehensive treatment and explanation. 

\subsubsection{\label{sec:lowenergies}Low energies}
As a final check we explore the low-energy limit of the amplitude to show that it agrees with the amplitude computed within the EFT of $\pi$. This EFT is obtained by integrating out the radial mode $h$ at tree level and at leading order in the derivative expansion. Starting from the equation of motion of $h$,
\begin{equation}
    \Box h -\frac{\mu^2 \dot{\pi}+(\partial \pi)^2}{v^2}(h+v)+\lambda(h^2 + 2 h v)(h+v)=0 \,,
\end{equation}
at leading order in the derivative expansion we have
\begin{equation}
    h^2+2hv=\frac{\mu^2 \dot{\pi}+(\partial \pi)^2}{\lambda v^2} \,.
\end{equation}
Plugging this expression into Eq.~\eqref{eq:Lagrangianexpanded} yields the low-energy Lagrangian of $\pi$ at leading order in the derivative expansion (neglecting a total derivative):
\begin{equation}
    \mathcal{L}_{\text{EFT}}=\frac{(\partial \pi)^2}{2}+\frac{(\mu^2 \dot{\pi}+(\partial \pi)^2)^2}{4\lambda v^4} \,.
\end{equation}
Canonically normalizing by setting $\pi = \pi_c c_s$ yields
\begin{equation}
\label{eq:cnEFT}
    \mathcal{L}_{\text{EFT}}=\frac{1}{2}\left(\dot{\pi}_c^2-c_s^2 (\boldsymbol{\nabla}\pi_c)^2\right)+\frac{\mu^2c_s^3}{2 \lambda v^4} \dot{\pi}_c(\partial \pi_c)^2+ \frac{c_s^4}{4 \lambda v^4}(\partial \pi_c)^4 \,,
\end{equation}
where
\begin{equation}
\label{eq:cs}
    c_s^2= \frac{1}{1+ \mu^4/2\lambda v^4}\equiv \frac{1}{1+\alpha} \,.
\end{equation}

The $S$-matrix for $\pi \pi \to \pi \pi$ can be computed easily and its explicit expression is reported in appendix~\ref{appSMatrixLowEnergy}.
We will now check that the low-energy limit of the full $S$-matrix \eqref{eq:SMatrix} agrees with the expression obtained within the EFT. As before we specialize to a kinematical configuration with all energies equal. In this configuration, the low energy limit of the $S$-matrix is realized by considering energies much smaller than the mass of the radial mode. Keeping only the leading order terms in the derivative expansion and making use of 
\be
Z^{\pi}_{-}(0) = c_s \,, ~ E_{-}(0)^2 - E_{+}(0)^2 = -\frac{M^2}{c_s^2} \,,
\ee
the amplitude contributions reduce to ($s \equiv (p_1+q_1)^2$, $t \equiv (q_2-q_1)^2$ and $u \equiv(p_2-q_1)^2$)
\begin{equation}\label{eq:52}
    \begin{aligned}
   \mathcal{M}_c  \to -\alpha^2 \frac{ c_s^4}{\mu^4}\omega ^2 \left(7 \alpha  \omega ^2-2 s+2 t+2 u\right) \,, 
\end{aligned}  
\end{equation}
\begin{equation}\label{eq:53}
    \begin{aligned}
   & \mathcal M_s \to  -\frac{16 c_s^4 \alpha^2 \omega^2}{\mu^4}\frac{(s+\alpha \omega^2)^2}{s+4\alpha \omega^2}+c_s^4  \frac{ \alpha   \left(\alpha  \omega ^2+s\right)^2}{\mu ^4 } \,,
\end{aligned}
\end{equation} 
\begin{equation}\label{eq:54}
    \begin{aligned}
    &\mathcal M_u \to c_s^4\frac{\alpha  \left(3 \alpha  \omega ^2+u\right)^2}{\mu ^4} \,,
\end{aligned}
\end{equation}
\begin{equation}\label{eq:55}
    \begin{aligned}
   & \mathcal M_t \to c_s^4\frac{\alpha  \left(3 \alpha  \omega ^2+t\right)^2}{\mu ^4} \,.
\end{aligned} 
\end{equation} 
As a consequence, they sum up to give
\begin{equation}\label{eq:56}
    \begin{aligned}
        \mathcal{M}_{\mathrm{low-energy}}=\frac{c_s^4
        \alpha  \left(s^2+t^2+u^2-4 \alpha   \omega ^2\right)}{\mu ^4}-\frac{16 c_s^4 \alpha^2 \omega^2}{\mu^4}\frac{(s+\alpha \omega^2)^2}{s+4\alpha \omega^2} \,,
    \end{aligned}
\end{equation}
in agreement with the EFT computation (see Eq. \eqref{EFTresultM} in appendix \ref{appSMatrixLowEnergy}).

\section{\label{secdecay}Decay rate of the Goldstone}
The states $-$ and $+$ are unstable, so they are not exact asymptotic states. Therefore, the $S$-matrix we calculated makes sense only if the typical time of interaction, which is of order $\omega^{-1}$, is much shorter than the lifetime of the states. In terms of the decay rate one needs
\begin{equation}
\label{eq:consist_check}
   \frac{\Gamma (\omega) }{\omega} \ll 1 \, . 
\end{equation}
Before delving into the estimates of the decay rate, it is important to notice that the model at hand is weakly coupled at all energies. As such, we expect all interactions to be suppressed by powers of the coupling constant $\lambda$:  also $\Gamma$ is thus $\lambda$-suppressed. Therefore we expect that, for $\lambda \ll 1$, the condition \eqref{eq:consist_check} is satisfied. We will now check this expectation computing the decay width in the two regimes $\omega \ll M$ and $\omega \gg M$, where $M$ is the mass of the radial mode $h$.    
\subsection{$\omega \ll M$}
The gap of the $+$ branch, $E_{+}(0)$, is always of the same order as $M$.
For $\omega \ll M$ the only allowed decay process with a 2-particle final state is $-\to --$ and in this low-energy regime we can rely on the EFT Lagrangian~\eqref{eq:cnEFT}: the decay is induced   by the operator
\begin{equation}\label{eq:cubic4decay}
\frac{\mu^2c_s^3}{2 \lambda v^4} \dot{\pi}_c(\partial \pi_c)^2 \,  .
\end{equation}
The ratio \eqref{eq:consist_check} can be estimated as
\begin{equation}
\label{eq:gammaIR}
    \frac{\Gamma}{\omega}(\omega\ll M)\sim  (1-c_s^{-2})^2\left(\frac{\omega}{M}\right)^4 \frac{\mu^4 \lambda^2}{M^4}\sim \lambda \left(\frac{\omega}{M}\right)^4\left(\frac{\lambda \mu^4}{M^4}\right)^3  \ll \lambda \,.
\end{equation}
We have used $1-c_s^{-2}\approx -\mu^4/\lambda v^4$, $M^2\geq \sqrt{\lambda} \mu^2$ and the low-energy condition $\omega\ll M$.
The kinematical suppression $(1-c_s^{-2})^2$ requires some explanation. For small frequency the dispersion relation is linear, so that the kinematics of the decay is collinear. The scalar product $p_1^\mu p_{2\; \mu}$, which arises from the operator \eqref{eq:cubic4decay}, is thus proportional to $1-c_s^{-2}$.

\subsection{$\omega\gg M$} To tackle the high-energy side of the problem we begin by performing a convenient change of basis. In the UV it is useful to use a field basis where renormalizability is manifest. In this basis we will not need to worry about cancellations among various terms and it is thus simpler to make estimates. Such a basis is easy to find: the Lagrangian \eqref{eq:1} we started with is manifestly renormalizable and we simply need to expand around a symmetry breaking solution\footnote{Another natural choice of fields would have been
\be
\Phi \equiv \phi_b(t) + e^{i \mu^2 t / 2v} \frac{\sigma_1 + i \sigma_2}{\sqrt{2}} \, .
\ee The Lagrangian written in this basis reads
\be
\mathcal L =  \frac{(\partial \sigma_1)^2}{2} + \frac{(\partial \sigma_2)^2}{2} - \frac{ \mu^2}{2 v} \big  (\dot \sigma_1 \sigma_2  - \dot \sigma_2  \sigma_1  \big) -  \frac{\lambda}{4} \big ( 2 v \sigma_1 + \sigma_1^2 + \sigma_2^2  \big )^2 \, .  
\ee}
\begin{equation}
\label{eq:phi12}
    \Phi = \phi_b(t) + \frac{\phi_1 + i \phi_2}{\sqrt{2}} \, , \qquad  \phi_b(t) \equiv \frac{v}{\sqrt{2}} e^{i \mu^2 t/2 v} \;.
\end{equation}
Notice that one is allowed to treat $\phi_b$ as approximately constant in the regime $\omega \gg M$: indeed the frequency of $\phi_b$ is of order $\mu^2/v$, which is at most of order $M$.
For our estimates we need to know how much these new fields interpolate the two asymptotic states. As already discussed at the end of section~\ref{seccanonical-quantisation} it is easy to see that, above the crossover scale $\omega_c$,  both $\phi_1$ and $\phi_2$ interpolate to order one for $+$ and $-$, i.e. 
\begin{equation}
\begin{gathered}
    |\braket{0| \phi_{1,2} | \pm }| = \mathcal{O}(1) \,.
    \end{gathered}
\end{equation}
(Using this in the regime $M \ll \omega \ll \omega_c$ will give an upper bound for the decay rate, as we will discuss.) By plugging \eqref{eq:phi12} into \eqref{eq:1}, setting $t=0$ and focusing on the cubic part of the Lagrangian we obtain 
\begin{equation}\label{eq:CubicL}
    \mathcal L_{(3)} \sim - \lambda v \phi_1^2 \phi_2-\lambda v \phi_1^3 \, . 
\end{equation}
As expected this operator is relevant and we do not need to worry about cancellations. We can simply write
\begin{equation}
\label{eq:gammaUV}
    \frac{\Gamma}{\omega} (\omega \gg M) \sim \lambda^2 \frac{v^2}{\omega^2} \sim \lambda \frac{M^2}{\omega^2}
    \,.
\end{equation}
As expected, the ratio between the interaction time and the decay lifetime of $-$ is always suppressed by $\lambda$. Notice that the two estimates \eqref{eq:gammaIR} and \eqref{eq:gammaUV} do not agree when evaluated at $\omega=M$. Indeed, approaching $\omega = M$ from the low and the high-energy regime gives respectively
\begin{equation}
 \frac{\Gamma}{\omega} (\omega \ll M) \to \lambda \left( \frac{\lambda \mu^4}{M^4}\right)^3\;,\quad \frac{\Gamma}{\omega} (\omega \gg M) \to \lambda \>.
\end{equation}
This difference is a consequence of the fact that \eqref{eq:gammaUV} actually yields an overestimate of the $\Gamma/\omega$ ratio within the window of energies $M \ll \omega \ll \omega_c$. Even if these estimates are enough to conclude that the $S$-matrix is well defined since the decay is always perturbative, i.e.~suppressed by $\lambda$, it is instructive to work out a finer estimate for $\Gamma/\omega$ in the intermediate regime $M\ll \omega \ll \omega_c$. We will do this in Appendix \ref{appdecay}.

The careful reader might wonder why we only paid attention to the $-\to --$ process among the possible decay channels. What about the decay into $-+$ and $++$?
The decay into $++$ is kinematically forbidden: by energy and momentum conservation we should have 
\begin{equation}
    p^2 = (p_1 +p_2)^2 \qquad \text{where } p^2 < 0 \,, \quad p_i^2 > 0 \,. 
\end{equation}
Given that the sum of two timelike future-directed four-momenta, see eq.~\eqref{eq:disprelation}, is always timelike future-directed, we see that this channel is kinematically prohibited. We can try to repeat the same argument for $-$ decaying into $-$ and $+$. In this case we have 
\begin{equation}
    (p - p_1)^2 = p_2^2 \,,
\end{equation}
where we have taken the first outgoing particle to be gapless. Generically now we cannot conclude anything because the sum of two spacelike vectors is not necessarily spacelike.  If we restrict our attention to $\omega_1,\omega_2 <\omega < \omega_c$, we can approximate the dispersion relation as linear and then 
\begin{equation}
    \begin{aligned}
        (p - p_1)^2 & = p^2 + p_1^2 - 2 p_1 \cdot p_2 \simeq (c_s^2-1) ( \boldsymbol{p}^2 + \boldsymbol{p}_1^2 ) - 2 ( c_s^2 |\boldsymbol{p}_1| |\boldsymbol{p}_2| - \boldsymbol{p}_1 \cdot \boldsymbol{p}_2 ) \\ 
        & < (c_s^2-1) ( \boldsymbol{p}^2 + \boldsymbol{p}_1^2 ) - 2 ( c_s^2 |\boldsymbol{p}_1| |\boldsymbol{p}_2| - |\boldsymbol{p}_1| |\boldsymbol{p}_2|  ) = (c_s^2 -1) (|\boldsymbol{p}_1| - |\boldsymbol{p}_2|)^2 < 0 \, . 
    \end{aligned}
\end{equation}
Since $p_2^2 > 0$ we conclude that for $\omega < \omega_c$ this decay does not happen and the only kinematically allowed process in this regime is $- \rightarrow - -$. The decay $-\to -+$ only happens for $\omega \gtrsim \omega_c$. In this range of energies, the cubic part of the Lagrangian is again given by \eqref{eq:CubicL}, leading to the same estimate for the $\Gamma/\omega$ ratio of the $-\to -+$ process:
\begin{equation}
\frac{\Gamma}{\omega} (\omega \gtrsim \omega_c) \sim \lambda \frac{M^2}{\omega^2} \ll \lambda \,.
\end{equation}

It is important to stress that the non-analyticities studied in this paper are not a consequence of the absence of absolutely stable asymptotic states. Indeed, one can send the coupling $\lambda \to 0$ while keeping $\mu^2/v$ fixed. In this limit, $\Gamma \to 0$ while the discontinuity in $\mathcal{M}/\lambda$ remains constant.

\section{\label{secconclusions}Conclusions}
We calculated the $S$-matrix for  Goldstone scattering in a complex $\lambda |\Phi|^4$ model in a state at finite charge, i.e.~with spontaneous breaking of Lorentz invariance. This is arguably the simplest model that gives a Lorentz-violating $S$-matrix that is well-defined at all energies, and it may be a good testing ground to discuss general properties of the $S$-matrix when Lorentz invariance is broken. A general lesson that emerged from this calculation is that the $S$-matrix does not enjoy the same analyticity properties as in the Lorentz invariant case: the relation between the $S$-matrix and correlation functions introduces non-analyticities which are a consequence of the breaking of boosts.

It seems thus quite difficult to obtain general positivity bounds using the $S$-matrix, at least in the most naive way. It would be worth exploring whether one gains something looking at the limit of small breaking $\mu \to 0$: this would correspond to studying the $S$-matrix on a {\em weak} Lorentz-breaking background, similarly to the analysis of the propagation of signals on a background, pioneered in \cite{Adams:2006sv}. Alternatively, one could abandon the $S$-matrix and use other objects which are well-defined both in the UV and in the IR: one example are conserved currents as discussed in \cite{Creminelli:2022onn}, where some bounds were already derived.

Perhaps the more general lesson is that we have a very limited understanding of the basic properties of QFT when Lorentz is spontaneously broken.

\section*{Acknowledgements}
It is a pleasure to thank B.~Salehian for useful discussions. MD and LS are partially supported by the SNSF grant 200021\_213120.

\appendix

\section{\label{app2LSZ}Second derivation of LSZ}
In this appendix we derive the LSZ reduction formula starting from the expression of creation/annihilation operators. The final formula will be different from Eq.~\eqref{eq:LSZpolology}, before going on-shell, although it gives of course the same $S$-matrix. The conclusion about the lack of analyticity will remain the same. 

We need to write the creation/annihilation operators in terms of the fields and their time derivative. These are given at $t=0$ by 
\be\label{eq:PhiLadders}
    \begin{aligned}
        &\int \di^3 \boldsymbol{x} \,e^{-i \boldsymbol{k}\cdot \boldsymbol{x}} \phi^a(0, \boldsymbol{x}) = \sum_{l = \pm} \frac{Z^a_l a_l+\bar{Z}^a_la^{\dagger}_l}{2 E_l} \,, \\
        &\int \di^3 \boldsymbol{x} \,e^{-i \boldsymbol{k}\cdot \boldsymbol{x}} \dot{\phi}^a(0, \boldsymbol{x}) = \sum_{l = \pm} -i\frac{Z^a_l a_l-\bar{Z}^a_la^{\dagger}_l}{2} \,,
    \end{aligned}
\ee
where $a \in \{h,\pi\}$. We define 
\be\label{eq:59}
Y^a_l\equiv \frac{Z^a_l}{2 E_l}\>,\quad X^a_l \equiv -\frac{i}{2}Z^a_l \,.
\ee
Multiplying the first and second expression of \eqref{eq:PhiLadders} respectively by $\bar{Y}^{-1}$ and $\bar{X}^{-1}$ we get
\be\label{eq:60}
    \begin{aligned}
        &\int \di^3 \boldsymbol{x} \;e^{-i \boldsymbol{k}\cdot \boldsymbol{x}} (\bar{Y}^{-1})^m_a\phi^a(0,\boldsymbol{x}) = a^{\dagger}_m + \sum_{l = \pm} (\bar{Y}^{-1})^m_aY^a_l a_l \,,\\
        &\int \di^3 \boldsymbol{x}\; e^{-i \boldsymbol{k}\cdot \boldsymbol{x}} (\bar{X}^{-1})^m_a\dot{\phi}^a(0,\boldsymbol{x})= a^{\dagger}_m + \sum_{l = \pm} (\bar{X}^{-1})^m_aX^a_l a_l \,.
    \end{aligned}
\ee
Subtracting the two expressions we arrive at
\be\label{eq:61}
    \begin{aligned}
        & \sum_{l = \pm} D^m_l a_l(\boldsymbol{k}) = \int \di^3 \boldsymbol{x} \;e^{-i \boldsymbol{k}\cdot \boldsymbol{x}} (\bar{Y}^{-1})^m_a\phi^a(0,\boldsymbol{x})-\int \di^3 \boldsymbol{x} \; e^{-i \boldsymbol{k}\cdot \boldsymbol{x}} (\bar{X}^{-1})^m_a\dot{\phi}^a(0,\boldsymbol{x}) \,,\\
        &  \sum_{l = \pm} \bar{D}^m_l a^{\dagger}_l(\boldsymbol{k}) = \int \di^3 \boldsymbol{x} \;e^{i \boldsymbol{k}\cdot \boldsymbol{x}} (Y^{-1})^m_a\phi^a(0,\boldsymbol{x})-\int \di^3 \boldsymbol{x} \; e^{i \boldsymbol{k}\cdot \boldsymbol{x}} (X^{-1})^m_a\dot{\phi}^a(0,\boldsymbol{x}) \,,
    \end{aligned}
\ee
where
\be\label{eq:62}
D^m_l= (\bar{Y}^{-1})^m_aY^a_l-(\bar{X}^{-1})^m_aX^a_l \,.
\ee
Finally we have the expression of creation/annihilation operators we were after (reintroducing the time dependence):
\be\label{eq:63}
    \begin{aligned}
        &  a_l = F^l_a \int \di^3 \boldsymbol{x} \; e^{i E_l(\boldsymbol{k})t-i \boldsymbol{k}\cdot \boldsymbol{x}} \phi^a -G^l_a \int \di^3 \boldsymbol{x} \; e^{i E_l(\boldsymbol{k})t-i \boldsymbol{k}\cdot \boldsymbol{x}}\dot{\phi}^a \,,\\
        & a^{\dagger}_l=\bar{F}^l_a \int \di^3 \boldsymbol{x} \; e^{-i E_l(\boldsymbol{k})t+i \boldsymbol{k}\cdot \boldsymbol{x}}\phi^a-\bar{G}^l_a \int \di^3 \boldsymbol{x} \; e^{-i E_l(\boldsymbol{k})t+i \boldsymbol{k}\cdot \boldsymbol{x}}\dot{\phi}^a \,,
    \end{aligned}
\ee
with
\be\label{eq:64}
F^l_a=(D^{-1})^l_m (\bar{Y}^{-1})^m_a\>, \quad G^l_a=(D^{-1})^l_m (\bar{X}^{-1})^m_a \,.
\ee
Notice that creation/annihilation operators are written as a linear combination of the two fields (and their derivative), while in the discussion in the main text we were using a single field to interpolate the asymptotic states.

We can now perform the LSZ reduction:
\be
    \begin{aligned}
        &\langle q_2,p_2|p_1,q_1\rangle=\langle q_2,p_2| a^{\dagger}_{-,\text{in}}(\boldsymbol{q}_1)|p_1\rangle=\langle q_2,p_2| (a^{\dagger}_{-,\text{in}}(\boldsymbol{q}_1)-a^{\dagger}_{-,\text{out}}(\boldsymbol{q}_1))|p_1\rangle \\
        &=- \left( \lim_{x^0\to + \infty} - \lim_{x^0\to - \infty} \right) \int \di^3 \boldsymbol{x} \; e^{-i E_{-}(\boldsymbol{q}_1)t+i \boldsymbol{q}_1\cdot \boldsymbol{x}}\langle \bar{F}^{-}_a\phi^a- \bar{G}^{-}_a\dot{\phi}^a\rangle \\
        &=-\int \di x^0 \; \partial_0  \int \di^3 \boldsymbol{x} \; e^{-i E_{-}(\boldsymbol{q}_1)t+i \boldsymbol{q}_1\cdot \boldsymbol{x}}\langle \bar{F}^{-}_a\phi^a- \bar{G}^{-}_a\dot{\phi}^a\rangle \\
        &=- \int \di^4 x \; e^{-i E_{-}(\boldsymbol{q}_1)t+i \boldsymbol{q}_1\cdot \boldsymbol{x}}(-i E_{-}(\boldsymbol{q}_1)\bar{F}^{-}_a+i E_{-}(\boldsymbol{q}_1)\bar{G}^{-}_a\partial_0 + \bar{F}^{-}_a\partial_0 - \bar{G}^{-}_a\partial_0^2)\langle \phi^a\rangle \\
        &= \int \di^4 x \; e^{-i q_{1,-}\cdot x}\mathcal{D}^{-}_a \langle q_2,p_2| \phi^a |p_1\rangle \,.
    \end{aligned}
\ee
The differential operator $\mathcal{D}^{-}_a$ achieves the LSZ reduction on each leg. Let us look at its expression in Fourier space:
\begin{equation}\label{eq:Amputator}
    \mathcal{D}^{-}_a=\frac{(\omega-E_{-})(E_{-}^2-\boldsymbol{k}^2-M^2)}{(\boldsymbol{k}^2+M^2)(E_{+}^2-E_{-}^2)\bar{Z}^{\pi}_{-}}\begin{bmatrix}
       i(\boldsymbol{k}^2 \omega +M^2 \omega +E_{-}E_{+}^2)\; , &(v/\mu^2)(E_{+}^2-\boldsymbol{k}^2-M^2)(\boldsymbol{k}^2+M^2+\omega  E_{-})
    \end{bmatrix}_a \,.
\end{equation}

First of all let us see that this expression reduces to the usual one in the Lorentz invariant limit, $\mu = 0$.
In this case, 
\be\label{eq:66}
Y^a_l\to \frac{\delta^a_l}{2 E_l} \,, ~ X^a_l\to \frac{-i}{2}\delta^a_l \,, ~ D^l_m\to 2\delta^l_m \,, ~  F^m_a\to  E_m \delta^m_a \,, ~ G^m_a\to -i \delta^m_a 
\ee
so that
\begin{equation}\label{eq:LIAmputator}
    \mathcal{D}^{-}_a \to i \delta^{-}_a(E^2_{-}+ \partial_0^2) \,.
\end{equation}
This can be easily read from Eq.~\eqref{eq:Amputator}. Indeed, switching off the breaking implies the following limits on the dispersion relations:
\be\label{eq:68}
E_{+}^2 \to \boldsymbol{k}^2+M^2\>,\quad E_{-}^2\to \boldsymbol{k}^2 \,,
\ee
so that 
\begin{equation}\label{eq:69}
    \mathcal{D}^{-}_a\to (-\omega+E_{-})\begin{bmatrix}
        i(\omega+E_{-})\,,&0
\end{bmatrix}=\begin{bmatrix}
        i(E_{-}^2-\omega^2) \,,&0
    \end{bmatrix}_a
\end{equation}
which is the same as Eq.~\eqref{eq:LIAmputator}.

We now wish to show that this LSZ procedure is equivalent to what was obtained in the main text, Eq.~\eqref{eq:LSZpolology}. In terms of the field propagator $\Delta_{ab}$ this amounts to proving that
\be\label{eq:AmputatorVsPropagator}
\lim_{\omega\to E_{-}}\mathcal{D}^{-}_a \Delta_{ba}=\lim_{\omega \to E_{-}}\frac{(\omega^2-E_{-}^2)}{i\bar{Z}^{\pi}_{-}}\Delta_{ b \pi} \,.
\ee
Using the explicit form of the propagator,
\be\label{eq:71}
\Delta_{ab}=\frac{i}{2E_l}\left(\frac{Z^a_l \bar{Z}^b_l}{\omega-E_l}-\frac{\bar{Z}^a_l Z^b_l}{\omega+E_l} \right) \,,
\ee
one sees that the RHS of Eq.~\eqref{eq:AmputatorVsPropagator} yields $Z^b_{-}$. It is straightforward to verify that 
\be\label{eq:72}
        \lim_{\omega\to E_{-}}\mathcal{D}^{-}_a \Delta_{\pi a}=Z^{\pi}_{-} \,, ~ 
    \lim_{\omega\to E_{-}}\mathcal{D}^{-}_{a} \Delta_{h a} = Z^{h}_{-} \,.
\ee
Notice that the above equality holds only in the on-shell limit $\omega \to E_{-}$. The two LSZ procedures give different results off-shell and thus define different functions of the complex momenta.

\section{\label{appFeynman}Feynman rules}
In this appendix we explicitly compute the scattering amplitude \eqref{eq:SMatrix}. The first step is to find the Feynman rules. Let's start with the propagators. As we have seen the quadratic Lagrangian reads
\begin{equation}\label{eq:73}
    \begin{aligned}
   \tilde{\mathcal L}_{(2)}  
= \frac{1}{2} \Phi_i (-k) \Pi_{ij}  ^{(2)}(k) \Phi_j (k) \,,
\end{aligned} 
\end{equation} 
where we have defined 
\begin{equation}\label{eq:74}
    \Phi_h (k) \equiv \tilde h_k \,, ~~ \Phi_{\pi} (k) \equiv  \tilde \pi_k ~ \text{ and } ~ \Pi_{ij}  ^{(2)}(k) \equiv  \left( \begin{array}{cc}
    k^2  & i \mu^2 \omega /v \\
    -i \mu^2 \omega /v  & k^2-M^2 
\end{array} \right)_{ij} \,.
\end{equation}  
The propagators then read
\begin{figure}[ht]
    \centering
    \includegraphics[width=0.18\textwidth]{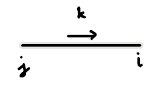}
\end{figure}
\vspace{-0.5cm}
\begin{equation}\label{eq:75}
    \begin{aligned}
     & = \frac{i}{(\omega^2 - E_{+}(\boldsymbol{k})^2) ( \omega^2 - E_{-}(\boldsymbol{k})^2)} \left( \begin{array}{cc}
    k^2-M^2  &  -i \mu^2 \omega /v \\
     i \mu^2 \omega /v  & k^2
\end{array} \right)_{ij} \\ 
& \equiv \Delta_{ij}(k) \,.
\end{aligned}
\end{equation}

We can now deal with the interacting Lagrangian. The cubic Lagrangian reads 
\begin{equation}\label{eq:76}
    \begin{aligned}
    \tilde{\mathcal L}_{(3) } & = - i \frac{\mu^2}{2 v^2} \omega_1 \tilde \pi_{k_1} \tilde h_{k_2} \tilde h_{k_3} - \frac{k_1 \cdot k_2}{v} \tilde \pi_{k_1} \tilde \pi_{k_2} \tilde h_{k_3} - \lambda v \, \tilde h_{k_1} \tilde h_{k_2} \tilde h_{k_3} \\ 
    & = - i \frac{\mu^2}{6 v^2} \left ( \omega_1 \tilde \pi_{k_1} \tilde h_{k_2} \tilde h_{k_3} + \omega_2 \tilde \pi_{k_2} \tilde h_{k_1} \tilde h_{k_3} + \omega_3 \tilde \pi_{k_3} \tilde h_{k_2} \tilde h_{k_1} \right) \\ 
    & - \frac{1}{3 v} \left ( k_1 \cdot k_2 \tilde \pi_{k_1} \tilde \pi_{k_2} \tilde h_{k_3} + k_1 \cdot k_3 \tilde \pi_{k_3} \tilde \pi_{k_1} \tilde h_{k_2} + k_3 \cdot k_2 \tilde \pi_{k_3} \tilde \pi_{k_2} \tilde h_{k_1} \right) \\ & -\lambda v \, \tilde h_{k_1} \tilde h_{k_2} \tilde h_{k_3}  \equiv \frac{1}{3!} \Pi^{(3)}_{i_1 i_2 i_3} (k_1,k_2,k_3) \Phi_{i_1}(k_1) \Phi_{i_2}(k_2) \Phi_{i_3}(k_3) \,.
\end{aligned} 
\end{equation}
Written explicitly the tensor $\Pi^{(3)}_{i_1 i_2 i_3} (k_1,k_2,k_3)$ reads 
\begin{equation}\label{eq:77}
    \begin{aligned}
&  \frac{1}{3!} \Pi^{(3)}_{i_1 i_2 i_3} (k_1,k_2,k_3)  = - \frac{i \mu^2}{v^2} \frac{ \omega_1 \delta_{i_1 \pi} \delta_{i_2 h}  \delta_{i_3 h} + \omega_2 \delta_{i_1 h} \delta_{i_2 \pi}  \delta_{i_3 h} + \omega_3 \delta_{i_1 h} \delta_{i_2 h}  \delta_{i_3 \pi}}{6} \\ & - \frac{k_1 \cdot k_2 \delta_{i_1 \pi} \delta_{i_2 \pi} \delta_{i_3 h} + k_1 \cdot k_3 \delta_{i_1 \pi} \delta_{i_2 h} \delta_{i_3 \pi} + k_2 \cdot k_3 \delta_{i_1 h} \delta_{i_2 \pi} \delta_{i_3 \pi}}{3v}  - \lambda v \, \delta_{i_1 h} \delta_{i_2 h} \delta_{i_3 h} \\ 
& = -\delta_{i_3 h} \left( \begin{array}{cc}
    \frac{k_1 \cdot k_2}{3 v}&  \frac{i \omega_2 \mu^2}{6 v^2}  \\
       \frac{i \omega_1 \mu^2}{6 v^2}& \lambda v
\end{array} \right)_{i_1 i_2} - \delta_{i_3 \pi} \left( \begin{array}{cc}
   0 & \frac{k_2 \cdot k_3}{3 v}  \\
      \frac{k_1 \cdot k_3}{3 v}& \frac{i \omega_3 \mu^2}{6 v^2}
\end{array} \right)_{i_1 i_2}. 
\end{aligned} 
\end{equation} 
The Feynman rule can be schematically written as 
\begin{figure}[ht]
    \centering
    \includegraphics[width=0.5\textwidth]{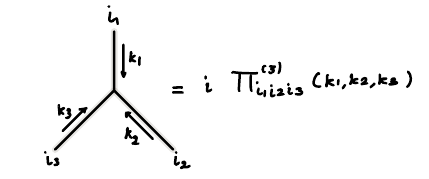}
\end{figure}

Finally we can turn to four point functions. The four-fields Lagrangian is 
\begin{equation}\label{eq:79}
    \begin{aligned}
    \tilde{\mathcal{L}}_{(4)} & = - \frac{k_1 \cdot k_2}{2 v^2} \tilde \pi_{k_1} \tilde \pi_{k_2} \tilde h_{k_3} \tilde h_{k_4} - \frac{\lambda}{4} \tilde h_{k_1} \tilde h_{k_2}\tilde h_{k_3} \tilde h_{k_4} \\ 
    & = - \frac{1}{ 12 v^2} \Big( k_1 \cdot k_2  \tilde \pi_{k_1} \tilde \pi_{k_2} \tilde h_{k_3} \tilde h_{k_4} + k_1 \cdot k_3  \tilde \pi_{k_1} \tilde \pi_{k_3} \tilde h_{k_2} \tilde h_{k_4} + k_1 \cdot k_4  \tilde \pi_{k_1} \tilde \pi_{k_4} \tilde h_{k_3} \tilde h_{k_2} \\  & + k_2 \cdot k_3  \tilde \pi_{k_2} \tilde \pi_{k_3} \tilde h_{k_1} \tilde h_{k_4} + k_2 \cdot k_4  \tilde \pi_{k_2} \tilde \pi_{k_4} \tilde h_{k_1} \tilde h_{k_3} + k_3 \cdot k_4  \tilde \pi_{k_3} \tilde \pi_{k_4} \tilde h_{k_1} \tilde h_{k_2} \Big) \\
    & - \frac{\lambda}{4} \tilde h_{k_1} \tilde h_{k_2}\tilde h_{k_3} \tilde h_{k_4} \equiv \frac{1}{4!} \Pi^{(4)}_{i_1 i_2 i_3 i_4} (k_1,k_2,k_3,k_4) \Phi_{i_1} \Phi_{i_2} \Phi_{i_3} \Phi_{i_4} \,.
\end{aligned} 
\end{equation} 
The Feynman rule reads 
\begin{figure}[ht]
    \centering
    \includegraphics[width=0.6\textwidth]{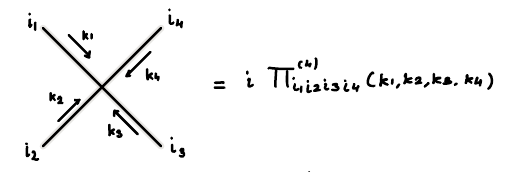}
\end{figure}

\newpage
\subsection{Scattering amplitudes}

\subsubsection*{Four-point}
\begin{figure}[h!]
    \centering
    \includegraphics[width=0.6\textwidth]{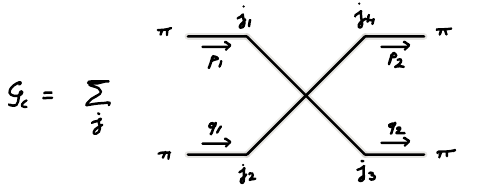}
\end{figure}

\begin{equation}\label{eq:81}
    \begin{aligned}
 = & \sum_{j} \Delta_{j_1 \pi } (p_1) \Delta_{j_2 \pi } (q_1) \Delta_{ \pi j_3 } (q_2) \Delta_{ \pi j_4 } (p_2) i \Pi^{(4)}_{j_1, j_2, j_3, j_4} (p_1, q_1, - q_2, - p_2) \\ & = \prod_{i=1}^2 \left (  \frac{i Z^\pi_{-} (\boldsymbol{p}_l)  }{(p^{0}_i)^2 - E_-(\boldsymbol{p}_l)^2} \frac{i Z^\pi_{-} (\boldsymbol{q}_l)  }{(q^{0}_i)^2 - E_-(\boldsymbol{q}_l)^2}  \right ) i \mathcal M_{\text{c}} \,,
\end{aligned}  
\end{equation} 
\begin{equation}\label{eq:82}
    \begin{aligned}
   &  \mathcal{M}_c = -  \frac{\mu^4}{v^4}  \prod_{i=1}^2 \left (  \frac{ Z^\pi_{-} (\boldsymbol{p}_i)^{-1}  }{(p^{0}_i)^2 - E_+(\boldsymbol{p}_i)^2} \frac{ Z^\pi_{-} (\boldsymbol{q}_i)^{-1}  }{(q^{0}_i)^2 - E_+(\boldsymbol{q}_i)^2}  \right )   \Big [6 \lambda \mu^4 p_1^0 q_1^0 q_2^0 p_2^0 \\ & - 2 p_1\cdot q_1 (p_1^2 - M^2) (q_1^2 - M^2) q_2^0 p_2^0 - 2 p_1\cdot p_2 (p_1^2 - M^2) (p_2^2 - M^2) q_1^0 q_2^0
   \\
   & - 2 p_1\cdot q_2 (p_1^2 - M^2) (q_2^2 - M^2) q_1^0 p_2^0 - 2 q_1\cdot p_2 (q_1^2 - M^2) (p_2^2 - M^2) p_1^0 q_2^0 \\ & - 2 q_2\cdot q_1 (q_2^2 - M^2) (q_1^2 - M^2) p_1^0 p_2^0 - 2 p_2\cdot q_2 (q_2^2 - M^2) (p_2^2 - M^2) p_1^0 q_1^0 \Big ] \,.
\end{aligned}  
\end{equation}

\subsubsection*{\texorpdfstring{$s$}{s}-channel}
\begin{figure}[ht]
    \centering
    \includegraphics[width=0.8\textwidth]{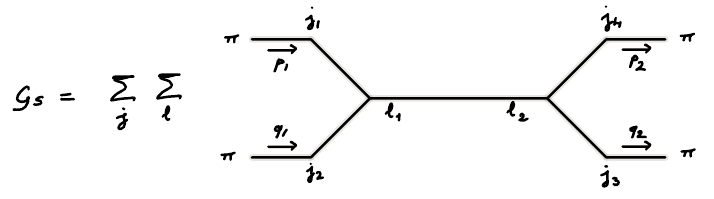}
\end{figure}

\begin{equation}\label{eq:83}
    \begin{aligned}
    & =\sum_{j} \sum_{l}  \Delta_{ j_1 \pi } (p_1) \Delta_{  j_2 \pi } (q_1) i \Pi^{(3)}_{j_1 j_2 l_1} (p_1,q_1,-p_1 - q_1) \Delta_{l_2 l_1} (p_1 + q_1) \\ & \times  \Pi^{(3)}_{l_2  j_3 j_4} (p_1 + q_1, - q_2, - p_2) \Delta_{\pi j_3} (q_2) \Delta_{\pi j_4} (p_2) \\ 
     & =  \sum_{l}  \Big [  \Delta  (-p_1)  i \Pi^{(3)}_{l_1} (p_1,q_1,-p_1 - q_1) \Delta (q_1) \Big ]_{ \pi \pi } \Delta_{l_2 l_1} (p_1 + q_1) \\ & \times \Big [ \Delta (p_1) i \Pi^{(3)}_{l_2 } (- p_1, - p_2,q_2 + p_2)  \Delta (-p_2) \Big ]_{ \pi \pi} \,.
\end{aligned}
\end{equation} 
So all together
\begin{equation}\label{eq:84}
    \begin{aligned}
    \mathcal{G}_s & = \Bigg [  v A(p_1,q_1) \left \{ v (p_1 + q_1)^2 A(p_1,q_1) - \mu^4 (p_1^0+q_1^0) B(q_2,p_2) \right \} \\ 
    & - \left \{ v \mu^4 B(p_1,q_1) \right \} \times \left \{ (p_1^0+q_1^0) A(q_2,p_2) + v (M^2 - (p_1 + q_1)^2) B(q_2,p_2) \right\} \Bigg ]
    \\  &  \prod_{i=1}^2 \left ( \frac{ 1}{p_j^2 (p_j^2 - M^2) - (\mu^4/v^2) (p_l^0)^2}\frac{ 1}{q_j^2 (q_j^2 - M^2) - (\mu^4/v^2) (q_l^0)^2} \right ) \\ & \times \frac{i v^{-10} }{(p_1 + q_1)^2 ( (p_1+q_1)^2 - M^2 ) - (\mu^4/v^2) (p_1^0+q_1^0)^2} \,,
\end{aligned}
\end{equation} 
where we have defined 
\begin{equation}\label{eq:85}
    A(p_1,q_1) \equiv 2 v^3 (M^2 - p_1^2) (M^2 - q_1^2) p_1 \cdot q_1 - v \mu^4 p_1^0q_1^0 (p_1^2 + q_1^2 + M^2) \,,
\end{equation}  
and 
\begin{equation}\label{eq:86}
    B(p_1,q_1) \equiv \mu^4 (p_1^0)^2 q_1^0  + 2 v^2 q_1^0 (M^2 - p_1^2) (p_1^2 + p_1 \cdot q_1) + (p_1 \leftrightarrow q_1) \,.
\end{equation}  
Finally we obtain
\begin{equation}\label{eq:87}
    \begin{aligned}
    \mathcal M_s & =  \prod_{i=1}^2 \left (  \frac{ Z^\pi_{-} (\boldsymbol{p}_i)^{-1}  }{(p^{0}_i)^2 - E_+(\boldsymbol{p}_i)^2} \frac{ Z^\pi_{-} (\boldsymbol{q}_i)^{-1}  }{(q^{0}_i)^2 - E_+(\boldsymbol{q}_i)^2}  \right ) \\ & \times \frac{ v^{-9}}{(p_1 + q_1)^2 ( (p_1+q_1)^2 - M^2 )  - (\mu^4/v^2) (p_1^0+q_1^0)^2} \\ 
    &  \Big [2 \lambda v^3 \mu^4  B(p_1,q_1) B(q_2,p_2) + [B(p_1,q_1) A(q_2,p_2) \\ & +B(q_2,p_2) A(p_1,q_1)] \mu^4 (p_1^0+q_1^0) - v [A(p_1,q_1) A(q_2,p_2) \\ & + \mu^4 B(p_1,q_1) B(q_2,p_2) ] (p_1 + q_1)^2   \Big ] \,.
\end{aligned}
\end{equation}

\subsubsection*{\texorpdfstring{$t$}{t}-channel}
\begin{figure}[ht]
    \centering
    \includegraphics[width=0.8\textwidth]{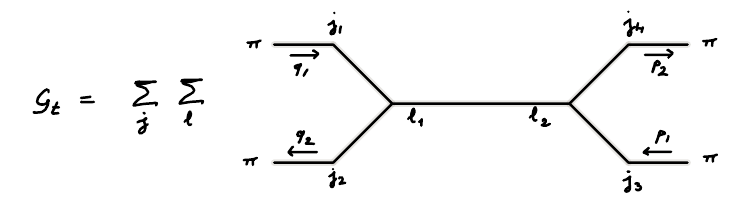}
\end{figure}

\begin{equation}\label{eq:88}
    \begin{aligned}
    = \sum_{j} \sum_{l} & \Delta_{ j_1 \pi } (q_1) \Delta_{  j_2 \pi } (-q_2) i \Pi^{(3)}_{j_1 j_2 l_1} (q_1,-q_2,-q_1 + q_2) \Delta_{l_2 l_1} (q_1 - q_2) \\ & \times  \Pi^{(3)}_{l_2  j_3 j_4} (q_1 -q_2,  p_1, - p_2) \Delta_{\pi j_3} (-p_1) \Delta_{\pi j_4} (p_2) \,.
\end{aligned} 
\end{equation} 
\begin{equation}\label{eq:89}
    \begin{aligned}
    \mathcal M_t & = \prod_{i=1}^2 \left (  \frac{ Z^\pi_{-} (\boldsymbol{p}_i)^{-1}  }{(p^{0}_i)^2 - E_+(\boldsymbol{p}_i)^2} \frac{ Z^\pi_{-} (\boldsymbol{q}_i)^{-1}  }{(q^{0}_i)^2 - E_+(\boldsymbol{q}_i)^2}  \right ) \times \\ & \frac{ v^{-9}}{(q_1 - q_2)^2 ( (q_1-q_2)^2 - M^2 ) - (\mu^4/v^2) (q_1^0 - q_2^0)^2} \\ 
    &  \Big[ 2 \lambda v^3 \mu^4  B(q_1,- q_2) B(-p_1,p_2) + [B(q_1,-q_2) A(-p_1,p_2) \\ & +B(-p_1,p_2) A(q_1,-q_2)] \mu^4 (q_1^0 - q_2^0) - v [A(q_1,-q_2) A(-p_1,p_2) \\ & + \mu^4 B(q_1,-q_2) B(-p_1,p_2) ] (q_1-q_2)^2   \Big ] \,.
\end{aligned} 
\end{equation}

\subsubsection*{\texorpdfstring{$u$}{u}-channel}
\begin{figure}[ht]
    \centering
    \includegraphics[width=0.8\textwidth]{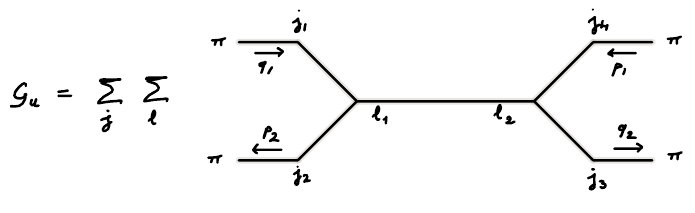}
\end{figure}

\begin{equation}\label{eq:90}
    \begin{aligned}
    = \sum_{j} \sum_{l} & \Delta_{ j_1 \pi } (q_1) \Delta_{  j_2 \pi } (-p_2) i \Pi^{(3)}_{j_1 j_2 l_1} (q_1,-p_2,-q_1 + p_2) \Delta_{l_2 l_1} (q_1 - p_2) \\ & \times  \Pi^{(3)}_{l_2  j_3 j_4} (q_1 -p_2, - q_2,  p_1) \Delta_{\pi j_3} (q_2) \Delta_{\pi j_4} (-p_1) \,.
\end{aligned} 
\end{equation}
Thus
\begin{equation}\label{eq:91}
    \begin{aligned}
    \mathcal M_u & = \prod_{i=1}^2 \left (  \frac{ Z^\pi_{-} (\boldsymbol{p}_i)^{-1}  }{(p^{0}_i)^2 - E_+(\boldsymbol{p}_i)^2} \frac{ Z^\pi_{-} (\boldsymbol{q}_i)^{-1}  }{(q^{0}_i)^2 - E_+(\boldsymbol{q}_i)^2}  \right ) \times \\  & \frac{ v^{-9}}{(q_1 - p_2)^2 ( (q_1-p_2)^2 - M^2 ) - (\mu^4/v^2) (q_1^0-p_2^0)^2} \\ 
    & \Big [2 \lambda v^3 \mu^4  B(q_1,-p_2) B(q_2,-p_1) + [B(q_1,-p_2) A(q_2,-p_1) \\ & +B(q_2,-p_1) A(q_1,-p_2)] \mu^4 (q_1^0-p_2^0) - v [A(q_1,-p_2) A(q_2,-p_1) \\ & + \mu^4 B(q_1,-p_2) B(q_2,-p_1) ] (q_1-p_2)^2   \Big ] \,.
\end{aligned}
\end{equation}

\section{\label{appSMatrixLowEnergy}\texorpdfstring{$S$}{S}-matrix in the low-energy EFT}
In this appendix, the EFT amplitude for the $\pi \pi \to \pi \pi$ process is reported. Making use of the Feynman rules coming from Lagrangian \eqref{eq:cnEFT}, the different contributions to the full amplitude can be evaluated:
\begin{displaymath}
    \begin{aligned}
        &\mathcal{M}_c=\frac{2c_s^4}{\lambda v^4}(p_1\cdot q_1 \, p_2 \cdot q_2 +q_1\cdot q_2 \, p_2 \cdot p_1+p_2\cdot q_1 \, p_1 \cdot q_2) \,, \\
        &\mathcal{M}_s=-\left( \frac{c_s^3 \mu^2}{\lambda v^4}\right)^2\frac{(2 (q_1^0+p_1^0)p_1 \cdot q_1+ p_1^0 q_1^2+q_1^0p_1^2)(2 (q_2^0+p_2^0)p_2 \cdot q_2+ p_2^0 q_2^2+q_2^0p_2^2)}{(q_1^0+p_1^0)^2-c_s^2 (\boldsymbol{q}_1+\boldsymbol{p}_1)^2} \,, \\
        &\mathcal{M}_t=-\left( \frac{c_s^3 \mu^2}{\lambda v^4}\right)^2\frac{(-2 (q_1^0-q_2^0)q_2 \cdot q_1- q_2^0 q_1^2+q_1^0q_2^2)(-2 (-p_1+p_2^0)p_2 \cdot p_1+ p_2^0 q_2^2-p_1^0p_2^2)}{(q_1^0-q_2^0)^2-c_s^2 (\boldsymbol{q}_1-\boldsymbol{q}_2)^2} \,, \\
        &\mathcal{M}_u=-\left( \frac{c_s^3 \mu^2}{\lambda v^4}\right)^2\frac{(-2 (q_1^0-p_2^0)p_2 \cdot q_1- p_2^0 q_1^2+q_1^0p_2^2)(-2 (q_2^0-p_1^0)p_1 \cdot q_2- p_1^0 q_2^2+q_2^0p_1^2)}{(q_1^0-p_2^0)^2-c_s^2 (\boldsymbol{q}_1-\boldsymbol{p}_2)^2} \,.
    \end{aligned}
\end{displaymath}
Specializing to equal energies $p_1^0=q_1^0=p_2^0=q_2^0=\omega$ yields a simpler expression for the $S$-matrix:
\begin{equation} \label{EFTresultM}
    \begin{aligned}
        & \mathcal{M}_c = \frac{2c_s^4}{\lambda v^4}(3 \alpha ^2 \omega ^4+\frac{s^2+u^2+t^2}{4}+\alpha  \omega ^2 (s+u+t))=\frac{c_s^4\alpha}{\mu^4}(-4 \alpha ^2 \omega ^4+s^2+u^2+t^2) \,, \\
        &\mathcal{M}_s=-\frac{16 c_s^4 \alpha^2 \omega^2}{\mu^4}\frac{(s+\alpha \omega^2)^2}{s+4\alpha \omega^2} \,, ~ \mathcal{M}_u=0 \,, ~ \mathcal{M}_t=0 \,.
    \end{aligned}
\end{equation}
    
\section{\label{appdecay}Estimate of the decay width}
In this appendix we wish to refine the estimate the decay width $\Gamma (\omega)$ for $\omega\gg M$. To this end, we work directly in the diagonal field basis $\varphi_{\pm}$ introduced in section \ref{secLSZ}. As we already pointed out there,
\begin{equation}
    \varphi_{\pm}(t,\boldsymbol{x}) \equiv  \int \frac{\di^3 \boldsymbol{k}}{(2\pi)^3 2 E_{\pm}(\boldsymbol{k})}\left(  a_{\pm}(\boldsymbol{k})e^{-i (E_{\pm}(\boldsymbol{k})t-\boldsymbol{k}\cdot \boldsymbol{x})}+ \text{h.c.} \right)
\end{equation}
are the field operators that interpolate only the two states $+$ and $-$ respectively. Recalling Eq. \eqref{eq:Phia} it is evident that $\pi$ and $h$ are linear combinations of the $\varphi_{\pm}$ fields. In particular, the $(h,\pi)$-basis is linked to the $\pm$ basis via the following transformation
\begin{equation}
    \begin{pmatrix}
        \tilde{\pi} ( k)\\
        \tilde{h} ( k)
    \end{pmatrix}
    = \begin{pmatrix}
        Z^{\pi}_{-}(\boldsymbol{k}) & Z^{\pi}_{+}(\boldsymbol{k})\\
         Z^{h}_{-}(\boldsymbol{k}) & Z^{h}_{+}(\boldsymbol{k})
    \end{pmatrix}
    \begin{pmatrix}
        \tilde{\varphi}_{-}( k )\\
        \tilde{\varphi}_{+} ( k)
    \end{pmatrix}\equiv \mathcal{U}(\boldsymbol{k})\begin{pmatrix}
        \tilde{\varphi}_{-} ( k)\\
        \tilde{\varphi}_{+} ( k)
    \end{pmatrix} \, . 
\end{equation}
The transformation $\mathcal{U}(\boldsymbol{k})$ that realizes the change of basis is a unitary matrix that diagonalizes the quadratic Lagrangian \eqref{eq:quadraticL}:
\begin{equation}
\begin{aligned}
     \mathcal{L}_{(2)} = \begin{pmatrix}
        \tilde{\varphi}_{-} ( - k )&\tilde{\varphi}_{+} ( - k )
    \end{pmatrix}\begin{pmatrix}
        k^2 - \frac{M^2}{2} + \frac{M^2}{2} \sqrt{ 1 + \frac{ 4  \omega^2}{\omega_c ^2}}&0\\
        0&k^2 - \frac{M^2}{2} - \frac{M^2}{2} \sqrt{ 1 + \frac{ 4  \omega^2}{\omega_c ^2}}
    \end{pmatrix}
    \begin{pmatrix}
        \tilde{\varphi}_{-} (k) \\\tilde{\varphi}_{+} (k) \end{pmatrix} \, . 
\end{aligned}
\end{equation}
Making use of the dispersion relations \eqref{eq:disprelation} we can trade the $\boldsymbol{k}^2$-dependence for $\omega$-dependence. For the interpolation functions we obtain
\begin{equation}
\label{eq:Zomega}
    Z^{\pi}_{-}(\omega)=Z^{h}_{+}(\omega)=\frac{\frac{1}{2}+\frac{1}{2}\sqrt{1+4\frac{\omega^2}{\omega_c^2}}}{\sqrt{2 \frac{\omega^2}{\omega_c^2}+\frac{1}{2}+\frac{1}{2}\sqrt{1+4\frac{\omega^2}{\omega_c^2}}}}\>,\>Z^{\pi}_{+}(\omega)=Z^{h}_{-}(\omega)=-i\frac{\omega/\omega_c}{\sqrt{2 \frac{\omega^2}{\omega_c^2}+\frac{1}{2}+\frac{1}{2}\sqrt{1+4\frac{\omega^2}{\omega_c^2}}}} \, . 
\end{equation}
Making use of this change of basis we can now rewrite the full Lagrangian in terms of the diagonal fields $\varphi_{\pm}$. Thus, if we are interested in the $- \to -- $ decay process we can simply focus on the trilinear vertices involving the $\varphi_{-}$ fields only. It is important to notice that, in general, the full Lagrangian will be non-local. The energy scale at which this non-locality will become important is $\omega_c \sim \lambda v^3/\mu^2$, introduced in Eq. \eqref{omegacEq}.  If we are interested in a given process at energies $\omega \ll \omega_c$, we can expand the interpolating functions 
\begin{equation}
\label{eq:Zlowomega}
    Z^{\pi}_{-}(\omega)=Z^{h}_{+}(\omega)\sim 1+ \mathcal{O}\left(\frac{\omega^2}{\omega_c^2}\right)\>,\>Z^{\pi}_{+}(\omega)=Z^{h}_{-}(\omega)\sim-i\frac{\omega}{\omega_c}+ \mathcal{O}\left(\frac{\omega^3}{\omega_c^3}\right) \, , 
\end{equation}
thus obtaining a local Lagrangian. Equivalently, to leading order in derivatives, we can write
\begin{equation}
    \pi \sim \varphi_{-}-i\frac{\dot{\varphi}_{+}}{\omega_c}\>,\>h\sim \varphi_{+}-i\frac{\dot{\varphi}_{-}}{\omega_c} \,.
\end{equation}
Plugging these expressions back into the cubic part of the Lagrangian and focusing on the trilinear vertices involving $\varphi_{-}$ only we obtain 
\begin{equation}
 \frac{(\partial \varphi_{-})^2 \dot{\varphi}_{-}}{v \omega_c} \, . 
 \end{equation}
By dimensional analysis we thus obtain
 \begin{equation}
      \frac{\Gamma}{\omega}(M\ll \omega \ll \omega_c)\sim  (1-c_s^{-2})^2\frac{\omega^4}{v^2 \omega_c^2}\sim \lambda \left( \frac{\omega}{\omega_c}\right)^4  \frac{\lambda \mu^4}{M^4}\ll \lambda \, . 
 \end{equation}
Again we have the same kinematical suppression factor discussed in section~\ref{secdecay}. This is a refinement of the previous overestimate \eqref{eq:gammaUV} and has the same functional form as the EFT decay width. We can now turn to the only missing case, namely $\omega \gg \omega_c$. The first idea that comes to mind is to expand the $Z$'s in $\omega_c/\omega$,
\begin{equation}
\label{eq:Zhighomega}
    Z^{\pi}_{-}(\omega)=Z^{h}_{+}(\omega)\sim \frac{1}{\sqrt{2}}+ \mathcal{O}\left(\frac{\omega_c}{\omega}\right)\>,\>Z^{\pi}_{+}(\omega)=Z^{h}_{-}(\omega)\sim-\frac{i}{\sqrt{2}}+ \mathcal{O}\left(\frac{\omega_c}{\omega}\right) \, , 
\end{equation}
to obtain a simplified Lagrangian suitable to compute processes in this regime. Although this looks technically very similar to what has been done just above, in essence it is rather different. In fact the $\omega_c/\omega$ expansion brings powers of energy to the denominator and as such yields non-local vertices, which are somewhat cumbersome to deal with. For these reasons we choose to work with the unexpanded Lagrangian instead:   
\begin{equation}
    \begin{gathered}
        \mathcal{L}_{(3)} \supset \Bigg( -\frac{i \omega_1 \mu^2}{2 v^2} Z^\pi_-(\omega_1) Z^h_-(\omega_2) Z^h_-(\omega_3) - \frac{k_1 \cdot k_2}{v} Z^\pi_-(\omega_1) Z^\pi_-(\omega_2) Z^h_-(\omega_3) \\ - \lambda v Z^h_-(\omega_1) Z^h_-(\omega_2) Z^h_-(\omega_3) \Bigg) \tilde \varphi_-(k_1) \tilde \varphi_-(k_2) \tilde \varphi_-(k_3) \,, 
    \end{gathered}
\end{equation}
and perform the expansion only at the end. The associated matrix element reads
\begin{equation}
    \begin{gathered}
        \mathcal M  = - \frac{i \mu^2}{v^2} \Bigg ( \omega_1 Z^{\pi}_{-}(\omega_1)Z^{h}_{-}(\omega_2)Z^{h}_{-}(\omega_3)+\omega_2 Z^{\pi}_{-}(\omega_2)Z^{h}_{-}(\omega_1)Z^{h}_{-}(\omega_3)+\omega_3 Z^{\pi}_{-}(\omega_3)Z^{h}_{-}(\omega_2)Z^{h}_{-}(\omega_1) \Bigg ) \\  -  \frac{2 }{v}\Bigg (p_1\cdot p_2 Z^{\pi}_{-}(\omega_1)Z^{\pi}_{-}(\omega_2)Z^{h}_{-}(\omega_3)+p_1\cdot p_3 Z^{\pi}_{-}(\omega_1)Z^{\pi}_{-}(\omega_3)Z^{h}_{-}(\omega_2) \\ +p_3\cdot p_2 Z^{\pi}_{-}(\omega_3)Z^{\pi}_{-}(\omega_2)Z^{h}_{-}(\omega_1)\Bigg) - 6 \lambda v \, Z^h_-(\omega_1) Z^h_-(\omega_2) Z^h_-(\omega_3) \, . 
    \end{gathered}
\end{equation}
In the high energy limit, using conservation of energy and momentum we find 
\begin{equation}
    \mathcal M = -i \frac{\mu^2}{\sqrt{2}v^2}(\omega_1-\omega_2) + i \frac{\mu^2}{\sqrt{2}v^2}(\omega_1-\omega_2)-i\frac{v \lambda}{\sqrt{2}} - i \frac{3 v \lambda}{\sqrt{2}} +\mathcal{O}\left(\frac{\omega_c}{\omega}\right) = - i 2 \sqrt{2} \lambda v + \mathcal{O}\left(\frac{\omega_c}{\omega}\right) \, . 
\end{equation}
By dimensional analysis we thus conclude that the ratio between the interaction time and the decay time for the $-\to --$ process can be estimated to be
\begin{equation}
    \frac{\Gamma}{\omega}(\omega \gg \omega_c)\sim \frac{ \lambda ^2 v^2 }{   \omega^2}\sim \lambda \frac{ M^2 }{   \omega^2}\ll \lambda \, ,
\end{equation}
in agreement with the result obtained in the $\phi_1,\phi_2$ field basis, Eq. \eqref{eq:gammaUV}. This is a further confirmation that the ratio between interaction time and the decay time is always suppressed by $\lambda$.

\newpage
\footnotesize
\bibliographystyle{klebphys2}
\bibliography{biblio.bib}
\end{document}